\documentclass[amsmath,amssymb,aps,prd,11pt,tightenlines,superscriptaddress,nofootinbib,preprintnumbers,notitlepage]{revtex4-1}

\usepackage{xcolor}
\usepackage[normalem]{ulem}
\usepackage{paralist}

\newcommand{\chisq}{\ensuremath{\chi^2}}
\newcommand{\ktmean}{\ensuremath{\langle k_T \rangle}}

\input{header}

\usepackage{lineno}
\usepackage{adjustbox}
\begin{document}

\title{Forward Neutrinos from Charm at Large Hadron Collider}

\author{Atri Bhattacharya}
\email{a.bhattacharya@uliege.be}
\affiliation{Space sciences, Technologies and Astrophysics Research (STAR) Institute, Université de Liège, Bât.~B5a, 4000 Liège, Belgium}

\author{Felix Kling}
\email{felix.kling@desy.de}
\affiliation{Deutsches Elektronen-Synchrotron DESY, Notkestr.~85, 22607 Hamburg, Germany}

\author{Ina Sarcevic}
\email{ina@physics.arizona.edu}
\affiliation{Department of Physics, University of Arizona, Tucson, AZ 85721}
\affiliation{Department of Astronomy and Steward Observatory, University of Arizona, 
Tucson, AZ 85721}

\author{Anna M. Stasto}
\email{ams52@psu.edu}
\affiliation{Department of Physics, Penn State University, University Park, PA 16802}

\begin{abstract}
The currently operating FASER experiment and the planned Forward Physics
Facility (FPF) will detect a large number of neutrinos produced in
proton-proton collisions at the LHC. In this work, we estimate neutrino
fluxes at these detectors from charm meson decays, which will be particularly
important for the $\nu_e$ and $\nu_\tau$ channels.   We make prediction using
both the next-to-leading order collinear factorization and the
$k_T$-factorization approaches to model the production of charm quarks as well
as different schemes to model their hadronization into charm hadrons. 
In particular, we emphasize that a sophisticated modeling of hadronization
involving beam remnants is needed for predictions at FASER and FPF due to the
sensitivity to the charm hadron production at low transverse momenta and very
forward rapidity. As example, we use the string fragmentation approach
implemented in \texttt{Pythia~8}. While both standard fragmentation functions
and \texttt{Pythia~8} are able to describe LHCb data, we find that
\texttt{Pythia~8} predicts significantly higher rate of high energy neutrinos,
highlighting the importance of using the correct hadronization model when
making predictions. 

\end{abstract}

\maketitle 
\renewcommand{\baselinestretch}{0.85}\normalsize
\clearpage
\tableofcontents
\renewcommand{\baselinestretch}{1.0}\normalsize

%%%%%%%%%%%%%%%%%%%%%%%%%%%%%%%%%%%%%%%%%%%%%%%%%%%
\clearpage
\section{Introduction}
%%%%%%%%%%%%%%%%%%%%%%%%%%%%%%%%%%%%%%%%%%%%%%%%%%%

% motivation for studying forward charm production
The forward production of charm quarks in high-energy proton-proton collisions
at the LHC provides an excellent probe of the strong interactions. In the far
forward region, corresponding to pseudorapidity $\eta \gtrsim 7$, which is
beyond the coverage of the main LHC detectors, this process is sensitive to
parton distribution functions at small momentum fractions $x \sim 10^{-7}$ and
at a scale $Q \sim m_c$. In this region of very small-$x$ and small-$Q^2$, that
is not accessible to the direct measurement at HERA~\cite{H1:2009pze},
deviations from the collinear factorization approach may be expected and novel
small-$x$ dynamics can occur, see e.g. Ref.~\cite{Martin:2003us}. In
particular, the non-linear contributions, which lead to saturation effects of
the gluon density are expected to play an important role~\cite{Gribov:1983ivg}.
In addition, the fragmentation functions needed to predict $D$-meson production
are not well known in this regime, as they are usually constrained in $e^+e^-$
collisions, while the hadronic environment of the proton-proton collisions
introduces some new dynamical features which may lead to the factorization
breaking. Future measurements of forward charm production will therefore
provide a unique opportunity to study and test different aspects of QCD in this
novel kinematic regime. 

% Icecube connection 
The study of forward charm production is also important in the context of large
neutrino telescopes, such as IceCube. Here, charmed hadrons can be produced in
cosmic ray collisions and their decay constitutes the source of prompt
atmospheric neutrinos and hence a background to the extra-galactic neutrino
signal~\cite{Bhattacharya:2015jpa, Gauld:2015yia, Garzelli:2015psa}. There are
currently large uncertainties on the associated production rate and flux,
underpinned by the lack of input data both at colliders as well as at neutrino
telescopes. Indeed, IceCube has not seen evidence of prompt atmospheric
neutrinos and only sets an upper limit on the corresponding
flux~\cite{IceCube:2020wum}. New input on forward charm production from the LHC
will also improve the predictions for prompt atmospheric neutrino production. 

% LHC neutrino experiments 
The distribution of charmed hadrons at the LHC have been measured in the
central region by ATLAS~\cite{ATLAS:2015igt}, CMS~\cite{CMS:2017qjw,
CMS:2021lab} and ALICE~\cite{ALICE:2017olh, ALICE:2019nxm} and in the forward
region by LHCb~\cite{LHCb:2013xam, LHCb:2015swx, LHCb:2016ikn}. Together, these
measurements cover the pseudorapidities $|\eta|<4.5$, while at higher values
charm production remains as yet unconstrained. This situation will soon change
due to a new set of far forward experiments which will be able to detect and
study neutrinos produced at the LHC. Many of these LHC neutrinos originate from
the decay of charmed hadrons, and hence a measurement of the neutrino spectrum
allows us to indirectly constrain forward charm production. The first two
experiments, FASER$\nu$~\cite{FASER:2019dxq, FASER:2020gpr} covering $\eta>8.9$
and SND@LHC~\cite{SHiP:2020sos, Ahdida:2750060} covering $7.2<\eta<8.7$, have
started their operation with the beginning of LHC Run~3 in summer 2022.
Together, they will detect about ten-thousand neutrino interactions. Larger
detectors with the ability to detect about a million neutrino interactions have
been proposed in the context of the Forward Physics Facility
(FPF)~\cite{MammenAbraham:2020hex, Anchordoqui:2021ghd, Feng:2022inv}, which
would operate during the HL-LHC era.  

% way forward 
In anticipation of first data from the LHC neutrino experiments, it is
important to have a dependable modeling of forward charm production and
reliable predictions for the resulting neutrino fluxes and their uncertainties.
Such estimates of the neutrino flux are needed as input for a variety of
planned measurements, for example that of the neutrino interaction cross
section. In addition, a comparison of theoretical predictions with the neutrino
flux measurements will then allow to constrain QCD parameters, such as the mass
of the charm quark, factorization and renormalization scales, parton
distributions at small-$x$, and the fragmentation of the charm into D-mesons.
In this paper we address these questions and study charm production at the LHC
using two different QCD approaches: the perturbative collinear approach at
next-to-leading order (NLO) and the $k_T$-factorization approach. In particular
we constrain our models using available data from LHCb and make predictions for
the LHC neutrino experiments. 

Our main focus is on investigating the sensitivity of our calculations to the
different modeling of the fragmentation of charm quarks into hadrons. This is
especially important since the charmed hadrons are produced at very forward
rapidity and at  low transverse momenta. In this region additional effects may
occur due to the  interactions with beam remnants, and thus the standard
fragmentation approach which is suitable for high transverse 
momenta may not be an applicable description in this kinematics.
We base our analysis of different fragmentation schemes on the two different
QCD models of charm pair production mentioned above to ascertain where the major
source(s) of uncertainties and model dependence lie.
In particular, our detailed analysis using the \texttt{Pythia} Monte Carlo
event generator to model the fragmentation with color reconnection shows
significant differences in the forward rapidity region with respect to the
calculations using different fragmentation functions from the literature.
This demonstrates that the forward particle production and  the  resulting
high energy neutrino flux
is particularly sensitive to the physics of fragmentation.

% outline
The paper is organized as follows. In \cref{sec:experiments} we review the
experimental setup of the LHC neutrino experiments. We then discuss the
modeling of charm production via the perturbative NLO calculation and the
$k_T$-factorization approach in  \cref{sec:charm}. Different approaches to
modeling of the fragmentation, including standard fragmentation function
approach and  \texttt{Pythia}, are discussed in \cref{sec:frag}. The main
results are presented in \cref{sec:result}. Finally in \cref{sec:conclusion} we
present our summary and conclusions.

%%%%%%%%%%%%%%%%%%%%%%%%%%%%%%%%%%%%%%%%%%%%%%%%%%%
\section{\label{sec:experiments}Forward Neutrino Experiments at the LHC}

%%%%%%%%%%%%%%%%%%%%%%%%%%%%%%%%%%%%%%%%%%%%%%%%%%%

% particle production 
The production of flavored hadrons has been extensively studied at all four
main LHC experiments. This data provides a crucial input, for example for the
modeling of high energy cosmic ray collisions and atmospheric neutrino fluxes.
However, the most energetic hadrons are typically produced in the far forward
direction, which lies outside of the coverage of the main LHC detectors. These
particles are particularly relevant for modeling of cosmic ray collisions,
since they carry a large fraction of the air showers energy and are also the
source of the most energetic atmospheric neutrinos. While there are some
measurements on far forward hadron production using additional LHC detectors,
for example on photons and neutrons from LHCf~\cite{LHCf:2017fnw,
LHCf:2020hjf}, no data exists so far on strange and charm hadrons. Such input
would, however, be desirable to address the cosmic ray muon
puzzle~\cite{Cazon:2020zhx, Anchordoqui:2022fpn} as well as to improve
predictions for prompt atmospheric neutrino flux at neutrino
telescopes~\cite{Bhattacharya:2015jpa,Gauld:2015yia,Garzelli:2015psa}. This
situation is changing with the start of the LHC neutrino experiments, which
will provide novel constraints on the far forward production of flavored
hadrons. 

% forward neutrino experiments
Already in the 1980s it was noticed that the LHC would produce a large number
of neutrinos through the decay of hadrons~\cite{DeRujula:1992sn}. Indeed, at
each collision point the LHC generates an intense and strongly collimated beam
of high-energy neutrinos along the beam collision axis. About 480~m downstream
from the ATLAS interaction point this neutrino beam passes through the TI12 and
TI18 tunnels, which housed the injector during the LEP era but remained empty
during the LHC era. These locations provide unique opportunities to access the
neutrino beam and study its properties. The first measurement illustrating the
potential was performed by the FASER collaboration, which reported the first
neutrino interaction candidates using a small pilot detector in
2021~\cite{FASER:2021mtu}. Following this proof of feasibility, two dedicated
detectors have been installed in these locations. Located in TI12 is the FASER
experiment~\cite{FASER:2018ceo, FASER:2018bac, FASER:2022hcn}. While it is
mainly designed to search for light long-lived particles predicted by models of
new physics~\cite{Feng:2017uoz, Feng:2017vli, Kling:2018wct, Feng:2018pew,
FASER:2018eoc}, it also contains a dedicated emulsion neutrino detector called
FASER$\nu$~\cite{FASER:2019dxq, FASER:2020gpr}. This detector is centered on
the beam collision axis and covers the pseudorapidity range $\eta > 8.9$.
Located in TI18 on the opposite site of ATLAS is SND@LHC~\cite{SHiP:2020sos,
Ahdida:2750060}, which also contains an emulsion target as well as additional
electronic components. Unlike FASER, it is positioned slightly off-axis and
covers $7.2 < \eta < 8.7$. Both detectors have the ability to distinguish
neutrinos of different flavors and measure their energies. With the start of
Run~3 of the LHC in summer 2022, both experiments have now started their
operation and recently reported the observation of the first collider
neutrinos~\cite{FASER:2023zcr, Albanese:2023ryj}. During LHC Run3, which is
expected to last until 2025, the two experiments are expected to detect about
ten thousand neutrino interactions with TeV scale energies. 

% FPF
Upgraded detectors to continue the LHC neutrino program are envisioned for the
HL-LHC era. These would be located in the FPF~\cite{MammenAbraham:2020hex,
Anchordoqui:2021ghd, Feng:2022inv}, which is a dedicated cavern to be
constructed 620~m downstream of ATLAS with the space to house a suite of
experiments. In particular, three dedicated neutrino detectors have been
proposed in this context: the emulsion based neutrino detector FASER$\nu$2, the
electronic neutrino detector AdvSND, and the liquid noble gas based neutrino
detector FLArE. Due to a tenfold increase in both target mass and luminosity
these detectors have the potential to see more than a million neutrino
interactions and study their properties in greater detail. 

% Flux
A first estimate of the neutrino flux has been provided in
Ref.~\cite{Kling:2021gos}, taking into account both the prompt flux component
from charm decays occurring the interaction point as well as a displaced
component from the decay of long-lived light hadrons occurring further
downstream  from the interaction point. It uses a variety of different Monte
Carlo event generators to model the production of hadrons at the LHC and
employs a dedicated fast simulation to model the propagation and decay of
long-lived hadrons when passing through the LHC beam pipe and magnetic fields.
The results show that a majority of muon neutrinos and electron neutrinos at
low energy originate from the displaced decay of light hadrons, while high
energy electron neutrinos and tau neutrinos are mainly produced in the prompt
decay of charmed hadrons. It was also noted that (i) there are large
differences between the Monte Carlo generator's predictions for the prompt
neutrino flux component of about an order of magnitude at high energies, and
(ii) most of the generators have not yet been tuned or validated for charm
production. More reliable predictions for forward charm production are needed. 

Unlike for light mesons, the forward production of charm quarks can be
described using perturbative QCD. This provides a different approach to obtain
predictions for the LHC neutrino flux, which also offers a deeper connection to
the underlying theory of QCD. Several recent studies have presented
perturbative calculations for forward charm production at the LHC and derived
corresponding predictions for the associated neutrino fluxes. In
Ref.~\cite{Bai:2020ukz}, the authors employed the collinear factorization
approach at NLO, supplemented by additional $k_T$-smearing and fragmentation
functions to account for hadronization effects. Subsequent work by the same
authors explored the associated PDF uncertainties~\cite{Bai:2021ira} and the
connection to astroparticle physics~\cite{Bai:2022xad}, also see
Refs.~\cite{Jeong:2020idg, Jeong:2021vqp, Bai:2022jcs}. In
Ref.~\cite{Maciula:2022lzk, Maciula:2020dxv}, the authors used the
$k_T$-factorization approach, both in the full and hybrid realization, with
fragmentation functions and a recombination model for hadronization. They also
investigated the impact of an additional intrinsic charm component on the
forward neutrino flux.

In the present analysis, we consider both of these perturbative QCD approaches
and we particularly focus on the modeling of fragmentation.
We present our predictions for the charm production at LHCb and forward
neutrino fluxes at FASER. In the following sections, we provide detailed
descriptions of the forward charm production modeling employing both QCD
approaches.

%%%%%%%%%%%%%%%%%%%%%%%%%%%%%%%%%%%%%%%%%%%%%%%%%%%
\section{\label{sec:charm}Charm Quark Production in Hadronic Collisions}
%%%%%%%%%%%%%%%%%%%%%%%%%%%%%%%%%%%%%%%%%%%%%%%%%%%

% outline
The standard routine for calculating the charmed hadron production cross
section $\sigma_H$ is to fold the hadronic charm quark cross section $\sigma_c$
with a fragmentation function $F_{c \to H}$ 
\begin{equation}
    \sigma_H \sim \sigma_c \otimes F_{c\to H} \; .
\end{equation}
In this section, we first focus on the perturbative calculation of charm quark
production in hadronic collisions. In particular, we will describe and utilize
two QCD approaches to calculate charm quark production: the NLO collinear
factorization formalism and the $k_T$-factorization formalism. A detailed
discussion of fragmentation into hadrons will then be provided in
\cref{sec:frag}. 

%-------------------------
\begin{figure}
    \centering
    \includegraphics[height=5.5cm]{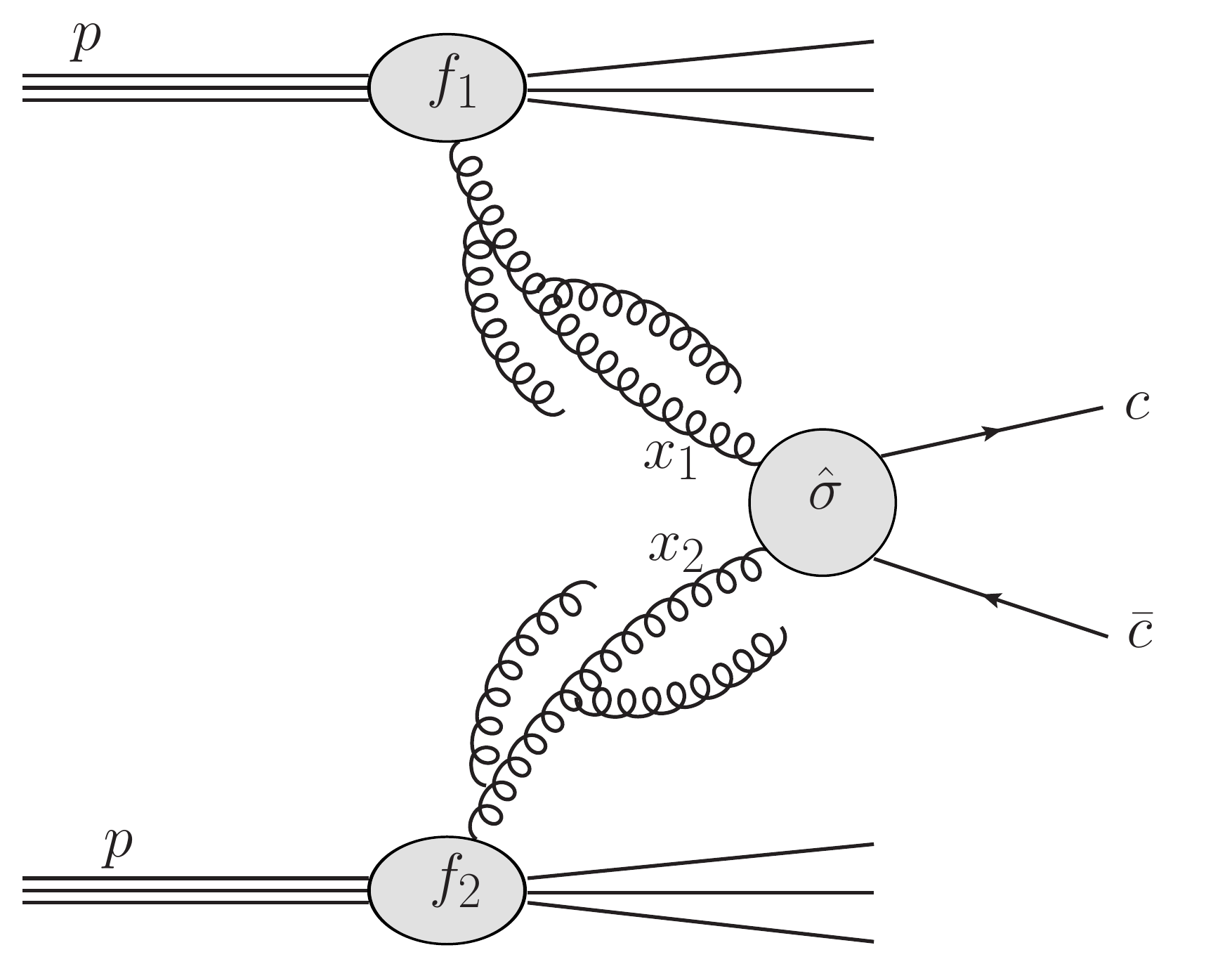}\hspace*{1.5cm}
        \includegraphics[height=5.5cm]{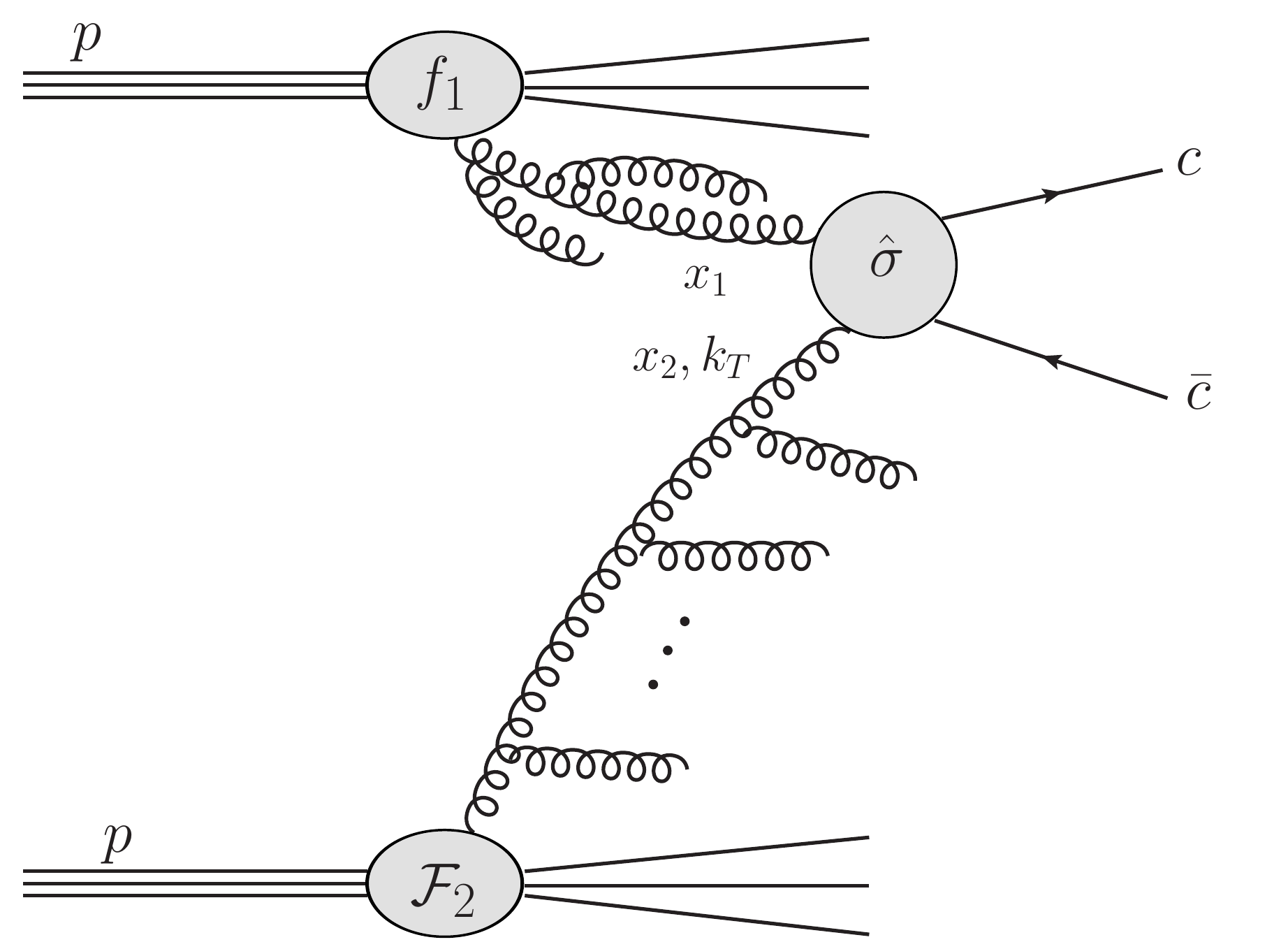}
    \caption{Left: gluon-gluon fusion process for charm production in
    hadron-hadron collisions in the collinear factorization approach. $f_1,f_2$
  are the integrated gluon distribution functions which depend on the
longitudinal momentum fractions $x_1,x_2$ and the hard scale of the partonic
sub-process.  Right: the same process, illustrated for the case of forward
production in the $k_T$-factorization. The gluon $x_1$ is treated on-shell, and
the gluon $x_2$ is off-shell with transverse momentum $k_T$. $\hat{\sigma}$ is
the partonic cross section which is on-shell (left panel) and takes into
account off-shellness of one gluon (right panel).}
    \label{fig:charmproduction}
\end{figure}
%-------------------------

The  production of charm in hadronic collisions is dominated by the gluon-gluon
scattering. In this process, gluons from two colliding hadrons fuse and produce
a charm quark-antiquark pair which subsequently fragments into the hadrons. The
generic diagram for gluon-gluon fusion process in hadronic collision is
illustrated in the left panel \cref{fig:charmproduction}, where the cross
section can be factorized into two gluon distribution functions, $f_1$ and
$f_2$, and the perturbatively calculable partonic cross section $\hat{\sigma}$.
This is the framework at the root of collinear factorization approach. 

In the forward region, this process probes the kinematics where the two
incoming partons have very different longitudinal momenta. The longitudinal
momentum of the forward charm quark at high energy is approximately equal to
$x_F \simeq E_c/E_p$ where $E_p$ is the energy of the incident proton and $E_c$
is the energy of the charm quark. Since we are interested in TeV-energy
neutrinos from TeV-energy charm decay, the corresponding forward charm
production kinematics probes values of $x_F$ of order $0.1$ or higher. This in
turn means that the longitudinal momentum fraction of  one of the gluons is
large, $x_1 \sim x_F$, and the other one  is very small. To be precise the
longitudinal momentum fraction $x_2 \simeq m_{c\bar{c}}^2 / (x_F s)$,  where
$m_{c\bar{c}}$ is  the invariant mass of the produced charm-quark pair and
$\sqrt{s}$ is the center of mass energy of the hadronic collision. This means
that for high energies the forward production is particularly sensitive to the
gluon density at very low values of $x_2 \gtrsim 10^{-7}$, which is not
constrained very well in this region. Thus the forward production offers unique
possibilities for tests of novel QCD dynamics in the region of small-$x$. 

%*************************************************
\subsection{\label{sec:nlo}Collinear Factorization at NLO}
%**************************************************

The double differential NLO cross-section for charm pair production is given by the expression
\begin{equation}
  \frac{d^2\sigma_{pp}}{dy dp^2_T}(s, m_c^2) =  \sum_{i,j = q, \overline q, g}
  \int dx_1 \, dx_2 \,  f_i (x_1,\mu_F^2) \,
  f_j(x_2,\mu_F^2) \, \frac{d^2\widehat{\sigma}_{ij}}{dy dp^2_T}(\hat{s},m_c^2,\mu_F^2, \mu_R^2)\,,
  \label{eq:nlo-xsec}
\end{equation}
where $m_c$ is the charm mass, $\surd{\hat{s}} = \sqrt{x_1 x_2 s}$ is the
partonic CM energy, $\mu_F$ and $\mu_R$ are the factorization and
renormalization scales respectively, and $f_{i,j}$ represent the quark and
gluon parton distribution functions (PDFs) as appropriate.  As noted
previously, we compute the cross-section to the next-to-leading order in
perturbation theory.

The double differential cross-sections for charm quark production are
calculated using the \texttt{FONLL} code~\cite{Cacciari:1998it,
Cacciari:2001td}, which provides an interface to
\texttt{LHAPDF}~\cite{Whalley:2005nh, Buckley:2014ana}, thus allowing one to
use a variety of up-to-date PDFs. We choose to use the central \texttt{CT14nlo}
PDF set~\cite{Dulat:2015mca} from the \texttt{LHAPDF} database as a
representative set for our analysis. While there are more recent PDF sets
available in the literature, including those that have been fit to 13 TeV LHCb
data and consequently have reduced uncertainties in their predictions at
low-$x$~\cite{Bertone:2018dse,Zenaiev:2019ktw}, we find that uncertainties in
the cross-section from scale variation dwarf those from using different PDFs.
Instead, our choice of the central \texttt{CT14nlo} PDF allows us to maintain
compatibility with results obtained in Ref.~\cite{Bhattacharya:2015jpa}, while
also using $m_c = 1.3$ GeV, consistent with the PDG best-fit.

To obtain best-fits to current charm data, we choose to vary the factorization
scale $\mu_F$ and renormalization scale $\mu_R$ while keeping the charm mass
fixed. Assuming the scales vary proportionally to the charm transverse mass,
$m_T = {\left( m_c^2 + p_T^2 \right)}^{1/2}$, it has been the norm to vary these parameters
independently within a range from $(0.5$--$2.0) \propto m_T$. However, when
restricting ourselves to this narrow range, we find that at high energies
$\sqrt{s} \geqslant 7$ TeV fits to data become progressively worse with
increasing rapidities. Furthermore, determining uncertainties around the
best-fit scales also requires one to extend the search beyond this range.
Therefore, we allow these parameters to vary independently over a broader range
$\in \left[ 0.5, 8.0 \right]$ unencumbered by theoretical preferences, allowing
the best-fit parameters to be instead determined by fitting to data. We also
determine the parameters defining a $1\sigma$ uncertainty band around the
best-fit cross-section.

We compute cross-sections for a range of parameters
$\left( \mu_R, \mu_F \right)$ and
obtain, for each choice of fragmentation scheme, the meson cross-section that may be
fit to data from LHCb.
The end result of this fitting exercise is that we obtain different sets of best-fit
$\left( \mu_R, \mu_F \right)$ for different fragmentation scheme.
We defer the details of our fitting procedure to \cref{sec:appendixa}.

%**************************************************
\subsection{\label{sec:kt}$k_T$-factorization} 
%**************************************************

In the forward  regime, one should apply a framework which incorporates
resummation of the large logarithms $\alpha_s \ln 1/x$. This is accomplished
through the $k_T$-factorization formalism~\cite{Catani:1990eg, Catani:1990xk,
Collins:1991ty}. The $k_T$-factorization formalism  involves off-shell
matrix-elements for partonic scattering and unintegrated gluon
distribution\footnote{In the context of the small-$x$ physics one traditionally
used the nomenclature of unintegrated parton distribution functions. There is
another formalism, see e.g.~\cite{Collins:2011zzd}, in which the corresponding
parton density functions are also transverse momentum dependent, they are
usually referred as TMDs. Relations between the two formalisms have been
extensively studied recently, see
e.g.~\cite{Kotko:2015ura},\cite{Balitsky:2015qba}.} functions ${\cal F}(x,{\bf
k}_T)$ which depend on the transverse momentum vector ${\bf k}_T$ of the
off-shell gluons. The unintegrated gluon distribution functions encode more
detailed information about the dynamics of the partons, and can be especially
important in providing information about the details of the kinematics of the
event. The $k_T$-factorization approach in hadroproduction of heavy quarks has
been considered in Refs.~\cite{Catani:1990eg, Catani:1990xk, Collins:1991ty}
where the off-shell matrix element  for heavy quark production have been
derived. The expression for the cross section in the $k_T$-factorization
formalism has the following form, see e.g.~\cite{Catani:1990eg}
\begin{equation}
      \sigma_{pp}(s,m_c^2) = 
  \int dx_1 \, dx_2 \, \frac{d^2{\bf k}_{1T}}{\pi} \, \frac{d^2{\bf k}_{2T}}{\pi} {\cal F} (x_1,{\bf k}_{1T}) \,
  {\cal F}(x_2,{\bf k}_{2T}) \, \widehat{\sigma}^{\rm off}(\hat{s},{\bf k}_{1T},{\bf k}_{2T},m_c)\,
  \label{eq:k_T-xsec}
\end{equation}
where the off-shell partonic cross section 
$\widehat{\sigma}^{\rm off}$
contains contributions from gluon-gluon scattering, dominant for the high
energy limit.
For the specific case of forward charm production considered here, due to the
fact the kinematics is very asymmetric and one gluon has  large longitudinal
momentum fraction $x_1$ it is appropriate to use an approach in which this
gluon is treated on-shell and satisfies the DGLAP evolution. Therefore the
formula \cref{eq:k_T-xsec} in this limit becomes
\begin{equation}
      \sigma_{pp}(s,m_c^2) = 
  \int dx_1 \, dx_2 \, \frac{d^2{\bf k}_{T}}{\pi} \,f (x_1,\mu^2) \,
  {\cal F}(x_2,{\bf k}_{T}) \, \widehat{\sigma}^{\rm on-off}(\hat{s},{\bf k}_T,m_c)\,,
  \label{eq:ybrid}
\end{equation}
where $\widehat{\sigma}^{\rm on-off}$ can be obtained from
$\widehat{\sigma}^{\rm off}$ by setting one gluon on-shell, see
Ref.~\cite{Catani:1990eg}. This is illustrated in the right panel of
\cref{fig:charmproduction}. The gluon with large longitudinal momentum fraction
$x_1$ is indicated together with schematically drawn collinear cascade
originating from one proton. On the other hand, the gluon with very small $x_2$
has transverse momentum $k_T$ and it is produced as a result of a very long
cascade of emissions from the other proton. These emissions are not collinear,
hence their transverse momenta are not ordered. Therefore such cascade leads to
the diffusion in the transverse momentum distribution. This approach was used
in Ref.~\cite{Martin:2003us} with the large $x_1$ gluon in the DGLAP collinear
regime, which is on-shell and the small $x_2$ gluon off-shell, with appropriate
approximation of the matrix element. 

In this work we are interested in the differential distributions in rapidity,
which can be obtained by generalizing collinear formula \cref{eq:nlo-xsec} to
include the transverse momentum dependence. Since we are using expressions
from~\cite{Catani:1990eg}, which are formally lowest order, the differential
cross section can be taken as 
\begin{equation}
    \frac{d\sigma}{dy_3 dy_4 d^2 {\bf p}_{3T} d^2 {\bf p}_{4T} } = \int
    \frac{d^2 {\bf k}_T}{\pi} \frac{\delta^{(2)}({\bf k}_T-{\bf p}_{3T}-{\bf
    p}_{4T})}{16 \pi^2 {(x_1 x_2 s)}^2} x_1 g(x_1,\mu^2) {\cal F}(x_2,{\bf k}_T)
    \overline{\sum} |{\cal M}^{\rm on-offshell}_{g g^*\rightarrow c\bar{c}}|^2
    \;,
\end{equation}
with momentum fractions
$x_1=\frac{m_{3T}}{\sqrt{s}}\exp(y_3)+\frac{m_{4T}}{\sqrt{s}}\exp(y_4)$ and
$x_2=\frac{m_{3T}}{\sqrt{s}}\exp(-y_3)+\frac{m_{4T}} {\sqrt{s}}\exp(-y_4)$ as
well as the transverse masses $m_{3,4T}^2=p_{3,4T}^2+m_c^2$ of the quark and
antiquark (see also~\cite{Maciula:2022lzk}).

The unintegrated gluon distribution functions within the high-energy formalism
need to be computed from the appropriate evolution equations which incorporate
the small-$x$ dynamics. The unintegrated parton densities within the high
energy formalism are usually computed from the Balitsky-Fadin-Kuraev-Lipatov
(BFKL) equation which resums the powers of $\alpha_s \ln
1/x$~\cite{Kuraev:1977fs, Balitsky:1978ic}. It has been computed at leading
logarithmic (LL) and next-to-leading logarithmic order (NLL)  in QCD. For the
phenomenological applications it needs to be supplemented by the additional
corrections which take into account higher orders in the form of kinematical
constraints and the constraints from matching to the DGLAP
evolution~\cite{Ciafaloni:2003rd}. In addition, in the limit of high energies,
or very small-$x$, other corrections are expected to occur, which are related
to the parton saturation phenomenon~\cite{Gribov:1983ivg}. In this regime, the
gluon densities are so large that recombination effects need to be taken into
account which are expected to slow down the growth of the gluon densities.
These corrections lead to the appearance of the non-linear terms in the
small-$x$ evolution equations. The non-linear evolution leads to the taming of
the gluon distribution in the region of very small-$x$ and moderate to small
values of scales $k_T$. To be specific, these evolution equations generate the
$x$-dependent saturation scale $Q_s^2(x)$. Whenever the relevant scale of the
process, say the $k_T$ of the gluon, is smaller than $Q_s^2(x)$ non-linear
terms are very important, while for $k_T^2 > Q_s^2(x)$ they can be neglected
and the non-linear evolution equations give results which coincide with thus
obtained form the linear evolution.

The effective theory for high density at small-$x$ is the Color Glass
Condensate~\cite{McLerran:1993ka, McLerran:1993ni, Jalilian-Marian:1997jhx,
Jalilian-Marian:1997qno, Iancu:2000hn, Ferreiro:2001qy}, with the corresponding
JIMWLK evolution equations. In the multicolor limit the hierarchy of JIMWLK
equations reduces to the Balitsky-Kovchegov equation~\cite{Balitsky:1995ub,
Kovchegov:1999yj}, the latter being the BFKL equation supplemented with the
nonlinear term in the gluon density.

The  small-$x$ unintegrated gluon density for the present paper was taken from
Ref.~\cite{Kutak:2012rf} as well as from Ref.~\cite{Li:2022avs}. The  gluon in
Ref.~\cite{Kutak:2012rf} which was based on the unified BFKL+DGLAP evolution
supplemented with small-$x$ resummation~\cite{Kwiecinski:1997ee}. Two sets of
gluon distributions were used: based on linear evolution as well as non-linear
evolution cast in the momentum space~\cite{Kutak:2003bd,Kutak:2004ym}. The
latter one includes the non-linear term in density which is responsible for the
saturation effects. Both sets of distributions were fitted to the data on $F_2$
structure function at HERA. The non-linear term is important for low-$x$ and
low values of transverse momenta and leads to taming of the gluon distribution
and therefore the resulting observable cross section.  We also used the gluon
extracted from more recent fit in Ref.~\cite{Li:2022avs} to HERA data, which
was based on the full resummation~\cite{Ciafaloni:2003rd,Ciafaloni:2003ek}
including the BFKL at NLO.

%%%%%%%%%%%%%%%%%%%%%%%%%%%%%%%%%%%%%%%%%%%%%%%%%%%
\section{\label{sec:frag}Charm Fragmentation}
%%%%%%%%%%%%%%%%%%%%%%%%%%%%%%%%%%%%%%%%%%%%%%%%%%%

% intro 
In the previous section we have discussed the perturbative aspects of charm
production. We now turn to question of fragmentation of charm quarks into charm
hadrons, which is a non-perturbative process and requires a separate treatment.
Here we first review the standard fragmentation function formalism as well as
its short-comings. We then present an alternative approach based on the
modeling of hadronization in Monte Carlo generators.

%**************************************************
\subsection{Fragmentation Functions} 
%**************************************************

% Fragmentation functions
Many studies of charm production at the LHC make use of the factorization
theorem to separate the charm production and fragmentation process. In the
literature, the latter is then modeled via fragmentation functions that have
been extracted from lepton collider data, assuming that they are also
applicable at hadron colliders. As we will explain later, this may not be
appropriate at hadron colliders, especially in forward and low transverse
momentum region that is most relevant for FASER. In this approach, one uses the
fact that charm quarks in electron-positron annihilation are produced with a
known momentum, for example with $p_c = m_Z/2$ at LEP. One can then measure the
flavor and momentum of charmed hadrons $p_H$ to constrain the fragmentation
process. This is typically parameterized in terms of fragmentation fractions
$f_H$, describing the probability of a charm quark to form a specific charm
hadron $H$, and a fragmentation function $D_H(z)$, describing the distribution
of fractional energy inherited by the hadrons $z=p_H/p_c$. In a later
comparison of fragmentation approaches, we use the fragmentation fractions
$f_{D^+} = 0.244$, $f_{D^0} = 0.606$, $f_{D_s^+} = 0.081$ and $f_{\Lambda_c^+}
= 0.061$ as obtained in Ref.~\cite{Lisovyi:2015uqa} and the \textit{Peterson}
fragmentation function~\cite{Peterson:1982ak}. It has the form $D_H(z) \sim
z^{-1}{[1-1/z-\epsilon/(1-z)]}^{-2}$ where we choose $\epsilon=0.035$ following
Ref.~\cite{ParticleDataGroup:2020ssz}. Note that the same fragmentation
function is used for all charmed hadrons. Simply for illustration, we will also
consider the unphysical case with \textit{no fragmentation} beyond
fragmentation fractions. This means that quark and hadron momenta are
identical, implying $D_H(z)=\delta(z-1)$.

% limitation of FF
Although the above-mentioned fragmentation functions approach has been
successfully applied to measurements of charm production in the central and
high-$p_T$ region of the LHC, it faces additional challenges in the forward and
low-$p_T$ regime. There are a variety of hadron collision measurements that
contradict the predictions obtained using fragmentation function; see
Sec.~6.2.2 of Ref.~\cite{Feng:2022inv} for a pedagogical overview. In the
following, we summarize three important observations that are particularly
relevant for the modeling of forward charm production:
\begin{itemize}
\item The first observation concerns the \textit{production asymmetry} of
  charmed mesons and their anti-particles. While the fragmentation function
  approach predicts equal production rates of charmed hadrons and their
  anti-particles, an excess of $D^-$ compared to $D^+$ has been observed at
  high $x_F$ in $\pi^-$-nucleus fixed target collisions recorded by
  WA82~\cite{WA82:1993ghz}, E769~\cite{E769:1993hmj} and
  E791~\cite{E791:1996htn}. Such production asymmetries in the forward
  direction are typically explained by charm hadronization involving the beam
  remnants~\cite{Norrbin:1998bw}. In the case of $\pi^-$-nucleus collisions,
  the $\bar c$ can hadronize with the valence $d$ from the pion and form an
  energetic $D^-$ meson. In contrast the formation of a $D^+$ requires a $c$
  and $\bar d$. Since the $\bar d$ cannot be a valence quark, but either a
  sea-quark or produced otherwise, the $D^+$ mesons are expected to be less
  energetic. This effectively induces a production asymmetry at high $x_F$.
\item The second observation regards the \textit{energy spectra}. Using the
  same data from pion fixed target experiments, it has been found that the
  momentum spectrum charm of hadrons are about as hard as or even harder than
  the charm quark spectra obtain from perturbation theory~\cite{WA82:1993ghz,
  E769:1993hmj, E791:1996htn}. This contradicts the fragmentation functions
  approach, which predict the hadrons to be softer than the charm quarks. In
  contrast, the above-mentioned mechanism of hadronization with other light
  quarks in the event, especially valence quarks from the beam remnant, would
  naturally allow the hadrons to be more energetic than the charm quarks and
  explain this observation. 
\item The third observation relates to the \textit{baryon to meson production
  ratios}. Recently, ALICE has measured the ratio between the $\Lambda_c$
  baryon and $D^0$ meson production rates in the central region and found that
  this ratio increases from about 10\% at high transverse momentum to about
  50\% at low transverse momentum~\cite{ALICE:2017thy, ALICE:2020wla,
  ALICE:2021rzj}. A similar enhancement was also seen by
  CMS~\cite{CMS:2019uws}. This disagrees with the expectation from
  fragmentation functions applied in the lab frame and extracted from LEP,
  which predict a roughly constant $\Lambda_c$ to $D^0$ ratio of around 10\%. 
\end{itemize}
The observations above illustrate that fragmentation functions extracted from
lepton colliders are not sufficient to describe charm production at hadron
colliders.

%**************************************************
\subsection{Hadronization using MC Generators} 
%**************************************************

% fragmentation in tools
One way to address the abovementioned problems is to use more sophisticated
models of fragmentation which are typically implemented in Monte Carlo
generators. Here, we will take advantage of these efforts and use
\texttt{Pythia~8}~\cite{Sjostrand:2006za, Sjostrand:2014zea} to model
hadronization. \texttt{Pythia} uses the Lund string
model~\cite{Andersson:1983ia, Sjostrand:1984ic} in which colored objects are
connected by a color string containing the field lines of the strong force.
This model can intuitively explain two of the above observations: a charm quark
connected to a beam remnant valence quark will be pulled forward, and hence
gain energy, or even hadronize together with the valence quark, leading to a
production asymmetry. By default, \texttt{Pythia} uses the \textit{Monash}
tune~\cite{Skands:2014pea}. While broadly used to describe phenomena at the
LHC, we note that it is not able to properly describe the baryon enhancement
observed at ALICE. This problem is addressed by a newer \textit{QCD-inspired
color reconnection} scheme introduced in Ref.~\cite{Christiansen:2015yqa}. It
allows for different string topologies, such as junctions of three strings,
which leads to a higher baryon production rates in high-multiplicity regions.
It has been also recently suggested~\cite{Kotko:2023ugv}, using modeling with
\texttt{Pythia}, that this QCD-inspired color reconnection mechanism might be
essential for the proper description of the $J/\psi$ production at the LHC.
Throughout this work, we use the \texttt{mode~2} configuration introduced in
Ref.~\cite{Christiansen:2015yqa}. 

One practical complication is that the tools we use to model the perturbative
production of charm quarks do not generate events that can be used as input to
\texttt{Pythia}, but only provide the charm quark distribution $d^2 \sigma_c /
(dp_{T,c} dy_c)$. We bypass this problem by using a re-weighting approach which
is inspired by Refs.~\cite{Gainer:2014bta, Mattelaer:2016gcx}. To understand
the underlying idea, let us recall that, conceptually, we can write the charm
hadron distribution $d^2\sigma_H / (dp_{T,H} dy_H)$ as a convolution of the
charm quark distribution $d^2 \sigma_c / (dp_{T,c} dy_c)$ and a (unitary)
transfer function $ f(\vec{p}_{c},\vec{p}_{H} )$ describing the hadronization
process:
\begin{equation}
  \frac{d^2\sigma_H}{dp_{T,H} dy_H} = \int \frac{d^2\sigma_c}{dp_{T,c}
  dy_c}  \times f(\vec{p}_{c},\vec{p}_{H} ) d \vec{p}_c. 
\label{eq:hadro}
\end{equation}
In general, the transfer function would depend on both the quark and hadron
momenta as well as the collider setup. In the fragmentation function approach,
we assumed that  $f = f_H \otimes D_H(z)$ and that it is independent of the
collider setup. In Monte Carlo generators, the hadronization procedure is more
complex and $f$ cannot be parameterized by a simple function. However, the
transfer function is encoded in a generated event output:  the charm production
process of \texttt{Pythia} provides a sample of events, where each event is
characterized by the parton momentum $\vec{p}_c$, the hadron momentum
$\vec{p}_H$, a hadron ID and an event weight $w$. The events in the sample
implicitly follow a distribution $d^2\sigma_c^{P8} / (dp_{T,c} dy_c)$ for the
charm quarks and $d^2\sigma_H^{P8} / (dp_{T,H} dy_H)$ for the charm hadrons
related via a \cref{eq:hadro} through a transfer function $f$.

To apply the same hadronization to a different model of charm production, we
use the re-weighting procedure and adjust the weights
\begin{equation}
    w \to w  \times \frac{d^2\sigma_c/(dp_{T,c}
    dy_c)}{d^2\sigma_c^{P8}/(dp_{T,c} dy_c)}. 
\end{equation}
By construction, the events will then follow a $d^2 \sigma_c / (dp_{T,c} dy_c)$
at quark level. The hadrons follow the desired distribution
\begin{equation}
   \int \frac{d^2\sigma^{P8}_c }{dp_{T,c} dy_c} \times
   \frac{d^2\sigma_c/(dp_{T,c} dy_c)}{d^2\sigma_c^{P8}/(dp_{T,c} dy_c)} \times
   f(\vec{p}_{c},\vec{p}_{H} ) \ d \vec{p}_c  =  \int
   \frac{d^2\sigma_c}{dp_{T,c} dy_c}  \times f(\vec{p}_{c},\vec{p}_{H} ) \  d
   \vec{p}_c  =
   \frac{d^2\sigma_H}{dp_{T,H} dy_H}  \ , 
\end{equation}
which we can extract from the event sample.

%Discussion 
Let us summarize our approach. The usual fragmentation function approach
assumes that the charm hadronization process is described by a transfer
function of the specific form $f=D_H(z)$, which is universal for all colliders,
applicable to all predictions of charm quark production, and independent of the
charm quark kinematics and hadronic environment. We saw, however, that this
assumption is invalid at hadron colliders. For example, hadronization with beam
remnants, that is not captured in the fragmentation functions, leads to a
harder forward charm hadron energy spectra and a charge asymmetry. This has
been observed at past beam dump experiments and is expected to be important for
forward charm hadron production at the LHC. 

We therefore propose an alternative approach to model charm hadronization using
\texttt{Pythia}, which only assumes that the underlying transfer function $f$
is the same for different predictions of charm quark production, and that
\texttt{Pythia} provides a reasonably good prediction of hadronization
especially in the forward direction. We note that the accuracy of
\texttt{Pythia}'s description of forward charm hadronization, especially with
beam remnants, has not yet been experimentally tested the hadronization process
due to a lack of experimental data. However, \texttt{Pythia}'s good description
of charm hadrons at beam dumps as well as light hadrons in the forward
direction of the LHC~\cite{LHCf:2017fnw, LHCf:2017fnw} provides some confidence
in its overall description of hadronization.

%%%%%%%%%%%%%%%%%%%%%%%%%%%%%%%%%%%%%%%%%%%%%%%%%%%
\section{\label{sec:result}Results}
%%%%%%%%%%%%%%%%%%%%%%%%%%%%%%%%%%%%%%%%%%%%%%%%%%%

% general intro
In this section, we present and discuss the results of our different charm
production models. We start this by systematically varying the modeling. For
each considered setup, we shall show comparisons of our predictions to the
double differential cross section of $D^0$ meson measured at 13~TeV by LHCb as
well as the expected neutrino event rates at FASER$\nu$. To determine the
neutrino flux, we follow the same approach as Ref.~\cite{Kling:2021gos}.
Initially, the charm hadrons are decayed in their rest frame according to the
decay branching fractions and energy distributions obtained with
\texttt{Pythia}. Subsequently, the resulting neutrinos are boosted into the
laboratory frame and recorded if they pass through the detector's
cross-sectional area. To obtain the anticipated number of neutrino interactions
in the target volume, we convolute the neutrino flux with the interaction
cross-sections obtained by \texttt{GENIE}~\cite{Andreopoulos:2009rq}. Here, we
consider FASER$\nu$ to consist of a $25~\cm \times 25~\cm \times 1~\m$ tungsten
target~\cite{FASER:2019dxq}.

In the following, we will present results for collinear factorization in
\cref{sec:results_nlo} and $k_T$-factorization in \cref{sec:results_kt}. We
will compare both approaches and show additional distributions in
\cref{sec:results_both}. 

%**************************************************
\subsection{\label{sec:results_nlo}Collinear Factorization at NLO}
%**************************************************

%-------------------------
\begin{figure}[t]
  \centering
  \includegraphics[width=0.99\textwidth]{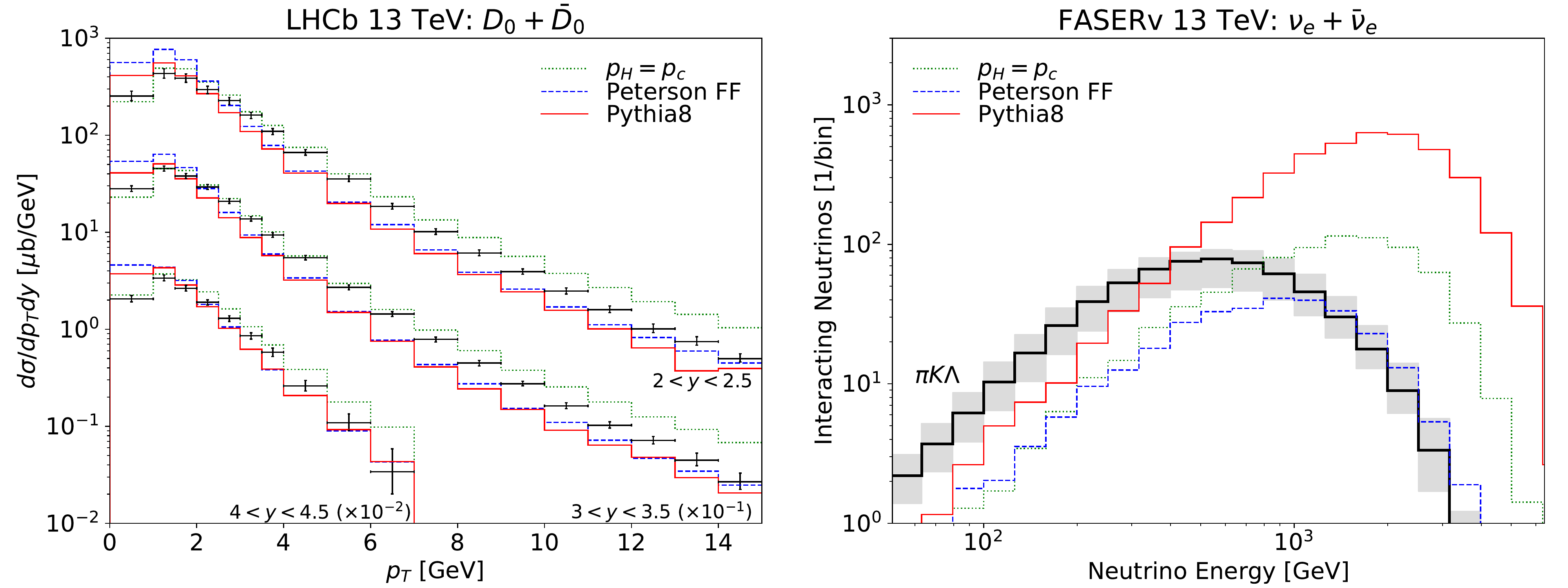}
  \caption{\label{fig:nlo_frag}\textbf{Modeling of Fragmentation:} Predictions obtained using
  collinear factorization at NLO using the \texttt{CT14nlo} parton distribution
functions. We show three different modeling approaches for fragmentation using
only fragmentation fractions (green dotted), using the Peterson fragmentation
function (blue dashed), and using \texttt{Pythia} with the QCD-inspired color
reconnection scheme (red solid). For each approach, the scales were obtained
using a fit to LHCb open charm data resulting in $(\mu_F,\mu_R)=(2.1,1.6)~m_T$
(no fragmentation function), $(\mu_F,\mu_R)=(3.75,1.75)~m_T$ (Peterson
fragmentation function) and $(\mu_F,\mu_R)=(2.25,1.5)~m_T$ (\texttt{Pythia}).
In the left panel, we compare these predictions with measurements of the double
differential neutral D-meson production rate obtained by LHCb at 13~TeV. We
present results for three different rapidity regions, where the results at
higher rapidity were scaled. In the right panel, we show the resulting number
of electron neutrinos from charm hadrons decay that interact with the
FASER$\nu$ detector as a function of the neutrino energy. For context, we also
display in black the event rate resulting from neutrinos from light hadron
decays as obtained in Ref.~\cite{Kling:2021gos}. See the main text for a
detailed discussion. }
\end{figure}
%-------------------------

\begin{figure}[t]
  \centering
  \includegraphics[width=0.99\textwidth]{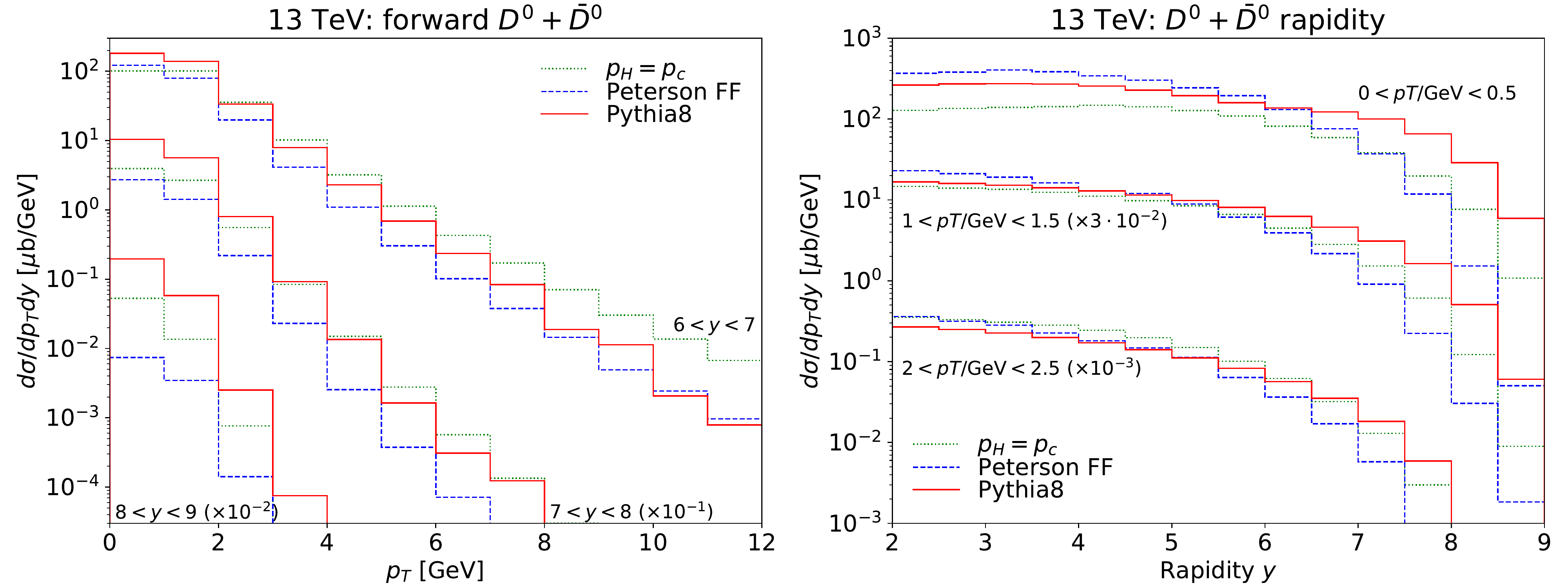}
  \caption{\label{fig:nlo_frag_2}\textbf{Predictions at High Rapidity:} Predictions obtained using
  the same QCD parameters and the same fragmentation functions as in Fig.2,
but for rapidity $y > 6$ (left panel) and the rapidity distributions for small
$p_T$ (right panel). }
\end{figure}
%-------------------------

% left panel, Fig 2
We first consider the calculation using the NLO collinear factorization. As
described in \cref{sec:nlo}, we obtain multiple best-fit cross-sections
corresponding to different fragmentation schemes. We find that a variation of
the scale parameters $(\mu_F, \mu_R)$ mainly influences the normalization of
the cross-section predictions, while the shape of the $p_T$ distribution
remains largely unchanged. In contrast, the latter is more significantly
affected by the choice of the fragmentation scheme.

We show a comparison of these results to the LHCb data in the left panel of
\cref{fig:nlo_frag} for three different modeling approaches for fragmentation.
The green dotted line shows the best fit prediction obtained using a constant
fragmentation factor. The best-fit cross-section in this case is obtained for
$(\mu_F, \mu_R) = (2.1, 1.6) \, m_T$ consistent with results
from~\cite{Bhattacharya:2015jpa}. However, we find that the $p_T$ shapes of the
corresponding double differential cross-sections are inconsistent with LHCb
data, consistently overestimating at high $p_T$. With change of scales
primarily affecting cross-section normalizations, and not the shape, there is
no way to improve the fit within the realm of our analysis when using constant
factors for fragmentation. Thus, this demonstrates the importance of including
more realistic fragmentation schemes. The blue dashed lines show the best fit
results using the Peterson fragmentation function, obtained for
$(\mu_F,\mu_R)=(3.75,1.75)\,m_T$. These agree reasonably well with LHCb data
for all rapidity regions, while still overestimating the data at low $p_T$
somewhat. Finally, best-fit results obtained using \texttt{Pythia} for
fragmentation are shown as red solid lines. These correspond to
$(\mu_F,\mu_R)=(2.25,1.5)\, m_T$. We observe that this setup produces similar
results to those using the Peterson fragmentation function in the regime
accessible to LHCb, with slight differences mainly at low $p_T < 2~\gev$.

We proceed to evaluate the electron neutrino flux from charm hadron decay at
FASER$\nu$ from these simulations. The results are shown in the right panel of
\cref{fig:nlo_frag}. With Peterson's fragmentation, the obtained flux has lower
rates and peaks at lower energies compared to the scenario without any
fragmentation. This outcome is expected since in the fragmentation function
approach, the charm hadron is always less energetic than the charm quark. In
contrast, using \texttt{Pythia} for fragmentation increases the neutrino flux
and shifts it to higher energies compared to the scenario without any
fragmentation. As discussed in \cref{sec:frag}, this outcome is consistent with
observations at beam dump experiments, where hadronization with beam remnants
plays a role. We emphasize that despite both fragmentation choices providing
similarly good descriptions of the LHCb data, they lead to a significant
difference in neutrino event rates at FASER$\nu$, differing by about one order
of magnitude. This highlights the importance of properly modeling fragmentation
for forward charm and, consequently, neutrino flux predictions for FASER$\nu$
and other LHC neutrino experiments. For comparison, we also show the event rate
from light hadron decays in black, as obtained in Ref.~\cite{Kling:2021gos},
using various generators. The solid line represents the central prediction,
while the shaded band shows the range of predictions from different generators.
This line is meant to provide optical guidance and to illustrate regions where
light and charm hadron decay contributions dominate the electron neutrino flux.

While our prediction already agrees  reasonably with the LHCb data, we observe
an underestimation of events at intermediate $p_T\sim8~\gev$ and a mild
overestimation at $p_T\sim1~\gev$ when compared to experimental measurements.
As pointed out in Ref.~\cite{Bai:2020ukz}, including an additional $k_T$
smearing, which aims to capture both an intrinsic transverse momentum of the
initial state partons as well as some soft gluon emission effects, can help
improve the agreement with data. The authors achieve this by introducing a
Gaussian smearing with width $\langle k_T \rangle$  of the transverse momentum
of the charm, while keeping its rapidity constant. However, we note that this
approach does not conserve energy and can lead to charm quarks that are more
energetic than the proton beam. Indeed, this leads to an unphysical order of
magnitude increase of the neutrino event rate at high energies. To address this
issue, we modify the smearing such that the $z$ component of charm quark
momentum is kept constant and the rapidity is allowed to change.

By iterating over a range of values of $\ktmean$ (see \cref{sec:app-nlo-ktsmear}),
we find that the best agreement to data is for a combination
of $\ktmean = 1.5~\gev$ and $(\mu_F,\mu_R) = (1.75,1.25)~m_T$.
This value of \ktmean\ is consistent with Ref.~\cite{Bai:2022xad}, which uses
$\ktmean = 1.2~\gev$, as well as with the default transverse
momentum for hard interactions used within \texttt{Pythia}, which is
$1.8~\gev$.
The corresponding neutrino fluxes are not highly sensitive to the choice
of \ktmean.

%-------------------------
\begin{figure}[tb]
  \centering
  \includegraphics[width=0.99\textwidth]{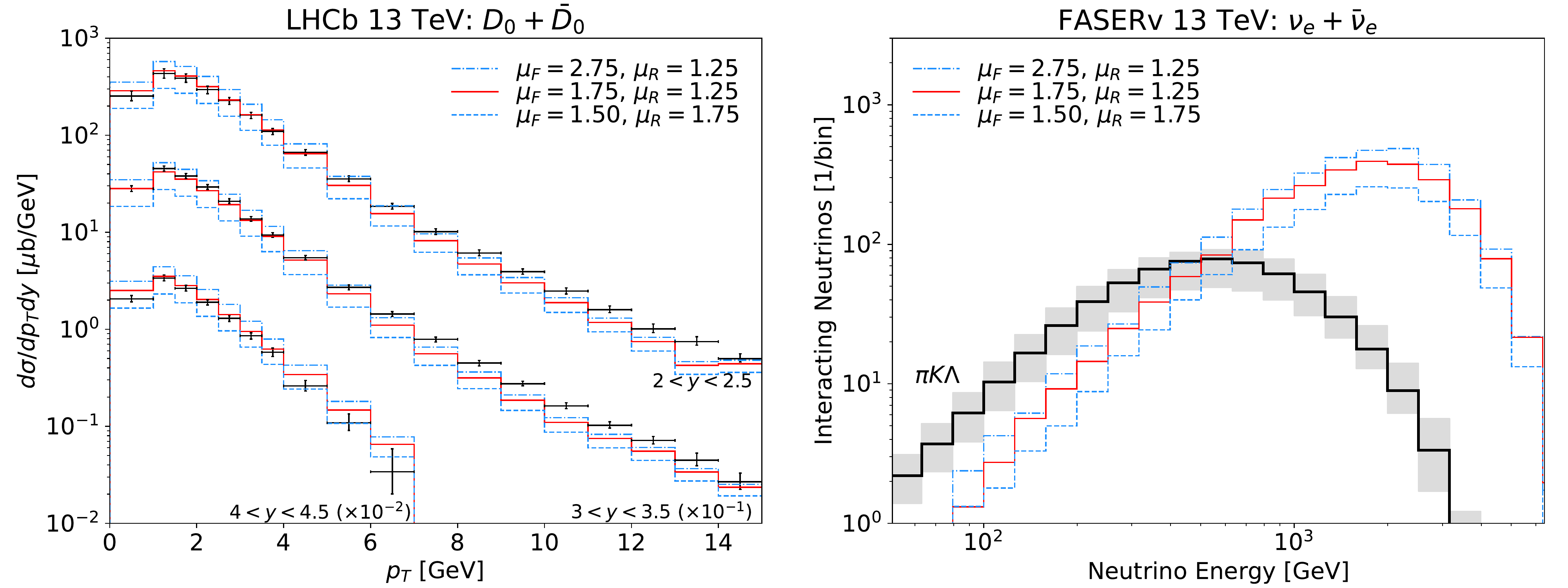}
  \caption{\label{fig:nlo_scales}\textbf{Scale Variation in Collinear
  Factorization:} Predictions using collinear factorization at NLO with
different choices of scales $\mu_F$ and $\mu_R$. All prediction use the
\texttt{CT14nlo} parton distribution function, $k_T$ smearing with $\langle k_T
\rangle = 1.5~\gev$ and \texttt{Pythia} with the QCD-inspired color
reconnection scheme to model fragmentation. See the main text for a detailed
discussion.}
\end{figure}
%-------------------------
In order to illustrate why different fragmentation approaches give similar
results for the LHCb data, but very different neutrino flux in the forward
region, we show $p_T$ distribution for large $y > 6 $ and 
the rapidity distributions for low $p_T$ in \figref{nlo_frag_2}. 
We note that for lowest value of $p_T$ and large rapidity, calculations with
various fragmentation schemes differ significantly.

Up to now, we've only shown our central prediction, which uses scale choices
$(\mu_F, \mu_R)=(1.75, 1.25)~m_T$ that were obtained by fitting the data with
$\langle k_T \rangle = 1.5~\gev$.
The same fit also allows to define scale uncertainties in a data driven way
(see \cref{sec:appendixa} for details).
To illustrate this, we present in \cref{fig:nlo_scales} our results for two
additional scale choices, which provide an error band that encompasses the LHCb
data.
Looking at the right panel, the corresponding neutrino fluxes show only mild
sensitivity to the choice of scales.  

%**************************************************
\subsection{\label{sec:results_kt}$k_T$-Factorization}
%**************************************************

We have observed that introducing an additional $k_T$ smearing improves the
agreement of the collinear factorization prediction with data. This smearing
effectively simulates intrinsic transverse momentum and soft-gluon emissions in
the initial state.  These effects are naturally included in the
$k_T$-factorization approach due to the presence of the unintegrated gluon
distribution function  and the off-shell matrix element which depend on
transverse momentum  $k_T$. As discussed in \cref{sec:kt}, we are using a
hybrid approach which utilizes an unintegrated PDF for the low-$x$ gluon and an
integrated PDF for the high-$x$ gluon.  This is because ultimately we are
interested in the very forward region where one $x$ is very small and the other
very large. 

%-------------------------
\begin{figure}[t]
  \centering
  \includegraphics[width=0.99\textwidth]{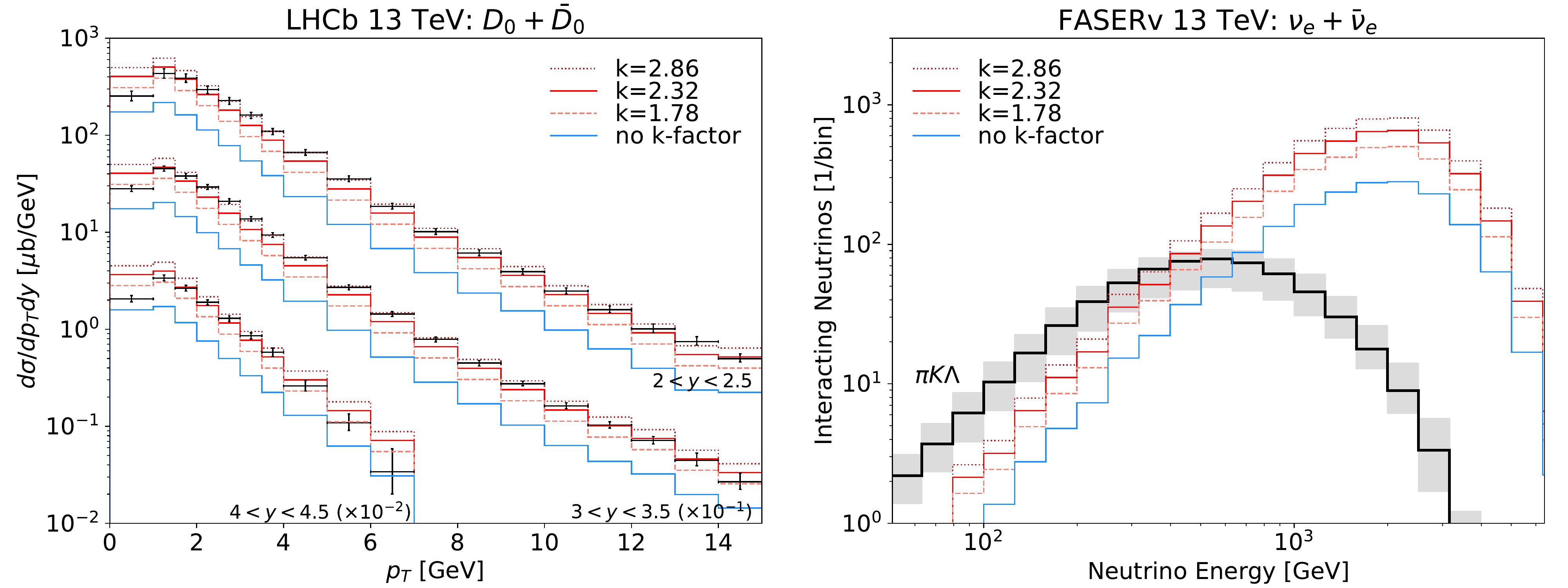}
  \caption{\label{fig:kt_norm}\textbf{Normalization in $k_T$-Factorization:} Predictions using
  $k_T$-factorization before (blue) and after (red) applying an overall
$k$-factor. These predictions use we use the \texttt{KS (non-linear)}
unintegrated distribution for the low-$x$ gluon, \texttt{CT14nlo} for the
high-$x$ gluon and use \texttt{Pythia} with the QCD-inspired color reconnection
scheme to model fragmentation. See the main text for a detailed discussion.}
\end{figure}
%-------------------------

As the basic setup we choose  the unintegrated gluon distribution from the
Kutak-Sapeta (\texttt{KS}) calculation~\cite{Kutak:2012rf} using the nonlinear
evolution, and for the large-$x$  we use the \texttt{CT14nlo} gluon.  Since the
\texttt{KS} gluon has been fitted to the HERA data using the leading order
strong coupling constant, we use the same setup for the one power of strong
coupling in the formula for the cross section. The second power of the coupling
is taken at NLO consistent with the \texttt{CT14nlo} PDF used for large-$x$
gluon. As before we are are modeling the  hadronization using \texttt{Pythia}
with the QCD-inspired color reconnection scheme. The results are shown in
\cref{fig:kt_norm} by the blue curve. We observe that the calculation has the
right shape in $p_T$ but it significantly underestimates the experimental data.
This was also observed in calculation of~\cite{Maciula:2022lzk}. This is not
totally unexpected since the off-shell partonic cross section used in
$k_T$-factorization is effectively computed at the LO~\cite{Catani:1990eg}.
Therefore when compared with NLO collinear calculation it does not have virtual
terms as well as final state gluon emissions from the quarks. It also has an
off-shell gluon only on the small-$x$ side. Given that the NLO calculation  in
the collinear approach resulted in $K$-factor of the order of $2.5$ with
respect to the LO result, see e.g.~\cite{Nason:1987xz}, it is expected that the
$k_T$-factorization will likely have large $K$-factor as well.

In order to get the normalization to agree with LHCb data, and therefore make
our extrapolations from LHCb to FASER$\nu$ more reliable, we introduce a
normalization factor which we refer to as $k$-factor\footnote{Traditionally a
$K$-factor refers to a ratio between the NLO and LO calculations. Since here we
are effectively using a normalization factor from lowest order to fit the data
we refer to it as $k$-factor to distinguish it from the one usually defined in
the literature.} in this calculation, determined by a fit to the data (with
additional weights that ensure each rapidity bin contributes identically to the
$\chi^2$ measure). We find a best fit of $k=2.32$; the resulting
double-differential cross-section  is illustrated by red line in
\cref{fig:kt_norm}. This is in excellent agreement with the LHCb data over the
full $p_T$ and rapidity range. This is encouraging since it means that the $x$
dependence of the unintegrated gluon, correctly reproduces the rapidity
dependence, and also the $p_T$ dependence is correctly captured. We shall also
see, that the $k$-factor does not change between the $7$ and $13$ TeV. We also
determine an uncertainty of the fit, as illustrated by the orange and magenta
curves in the same figure (using a rescaled $\chi^2$ for this following the PDG
procedure described in Refs.~\cite{ParticleDataGroup:2020ssz,
Rosenfeld:1975fy}). These variations form a nice envelope around the data with
a width  of about a factor 2 at low values of $p_T$. The right panel in
\cref{fig:kt_norm} shows the electron neutrino flux obtained in this approach.
A similar size band is also obtained at FASER$\nu$, see right panel.

%-------------------------
\begin{figure}[t]
  \centering
  \includegraphics[width=0.99\textwidth]{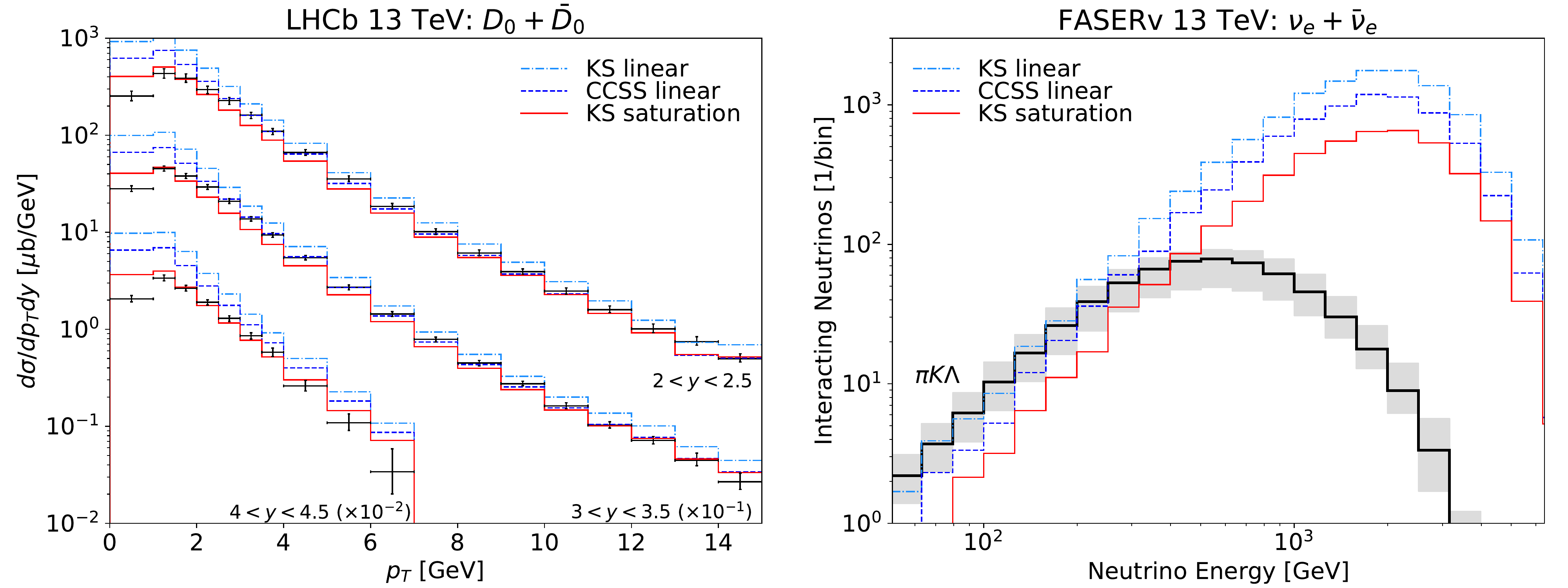}
  \caption{\label{fig:kt_lowx}\textbf{Low-$x$ Gluon Distribution in $k_T$-Factorization:}
  Predictions using $k_T$-factorization using the \texttt{KS} (blue dot-dashed)
and \texttt{CCSS} (blue dashed) unintegrated distribution with a purely linear
evolution as well as the \texttt{KS} unintegrated distribution including
non-linear effects that describe saturation effects (red solid) for the low-$x$
gluon. All predictions use a constant $k$-factor of $2.32$, the
\texttt{CT14nlo} parton distribution function for the high-$x$ gluon, use
\texttt{Pythia} with the QCD-inspired color reconnection scheme to model
fragmentation. See the main text for a detailed discussion.}
\end{figure}
%-------------------------

%low-x PDF
Next, we study the dependence of the results on the choice of the low-$x$
unintegrated gluon distribution. In \cref{fig:kt_lowx} we show our results for
three choices of unintegrated PDFs: two choices for the \texttt{KS} gluon with
linear evolution and with non-linear effects that describe saturation effects,
and  third choice of gluon from~\cite{Li:2022avs} obtained  from the linear
evolution including the  resummation using the Ciafaloni-Colferai-Salam-Stasto
(\texttt{CCSS}) approach~\cite{Ciafaloni:2003rd}.  We find that the prediction
which includes saturation effects is in excellent agreement with the LHCb data
over the full $p_T$ range. In contrast, the linear cases overshoot  the data at
low $p_T$. However, given that the results include the $k$-factor effectively
added by fitting as explained before, it is not possible to conclude at this
moment about the importance of the saturation effects in the LHCb data. 
Looking at the right panel, including saturation effects results in a reduction
of the flux by a factor of approximately three compared to the linear case.
This is due to the fact that the nonlinear effects are largest at very low
$p_T$.

We have also tested the sensitivity of the $k_T$ factorization calculations to
the choices of the large $x$ gluon distribution, the running coupling order and
the scale choice. The results of these studies are collected in
\cref{appendixb}. We have found rather small differences between the
calculations for these various choices.

%**************************************************
\subsection{\label{sec:results_both}Comparison of Approaches}
%**************************************************

Based on the previous discussion, we  identify central predictions for both
factorization approaches. In particular, we consider the following
configuration 
\begin{itemize}
    \item \textbf{collinear factorization at NLO} with \texttt{CT14nlo} for the
      gluon parton distribution, renormalization scale $\mu_R=1.75~\m_T$,
      factorization scale $\mu_F=1.25~m_T$, a $k_T$ smearing with $\langle k_T
      \rangle=1.5~\gev$, and fragmentation modeled with \texttt{Pythia} with
      the QCD-inspired color reconnection scheme
    \item \textbf{$k_T$-factorization} using \texttt{KS} unintegrated
      distribution for the low-$x$ gluon including saturation effects, the
      \texttt{CT14nlo} parton distribution for the high-$x$ gluon, a $k$-factor
      of $2.32$ and fragmentation modeled with \texttt{Pythia} with the
      QCD-inspired color reconnection scheme
\end{itemize}

%-------------------------
\begin{figure}[t]
  \centering
  \includegraphics[width=0.99\textwidth]{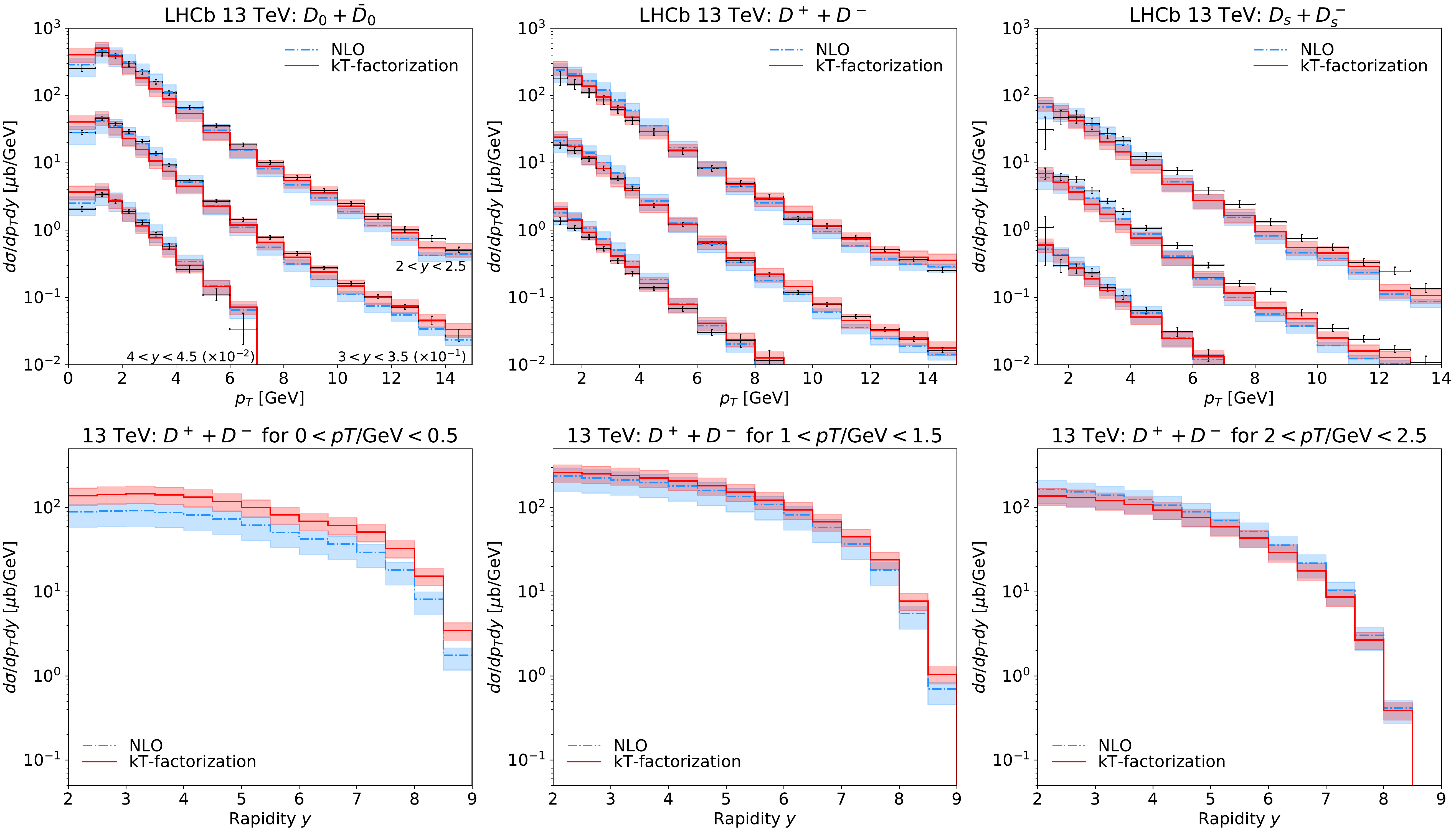}
  \caption{\label{fig:results1}\textbf{Comparison of Charm Hadron Distribution at 13 TeV:}
  Predictions using collinear factorization at NLO and $k_T$-factorization. The
shaded band around the NLO predictions corresponds to the scale variations
shown in \cref{fig:nlo_scales} while the shaded band around the
$k_T$-factorization prediction corresponds to a varation of the $k$-factor as
shown in \cref{fig:kt_norm}. In the top row, we show the $p_T$ distributions
for all three charmed mesons in comparison to LHCb data. The bottom row show
the rapidity distribution for $D^\pm$ mesons in three transverse momentum
regions.}
\end{figure}
%-------------------------

%-------------------------
\begin{figure}[t]
  \centering
  \includegraphics[width=0.99\textwidth]{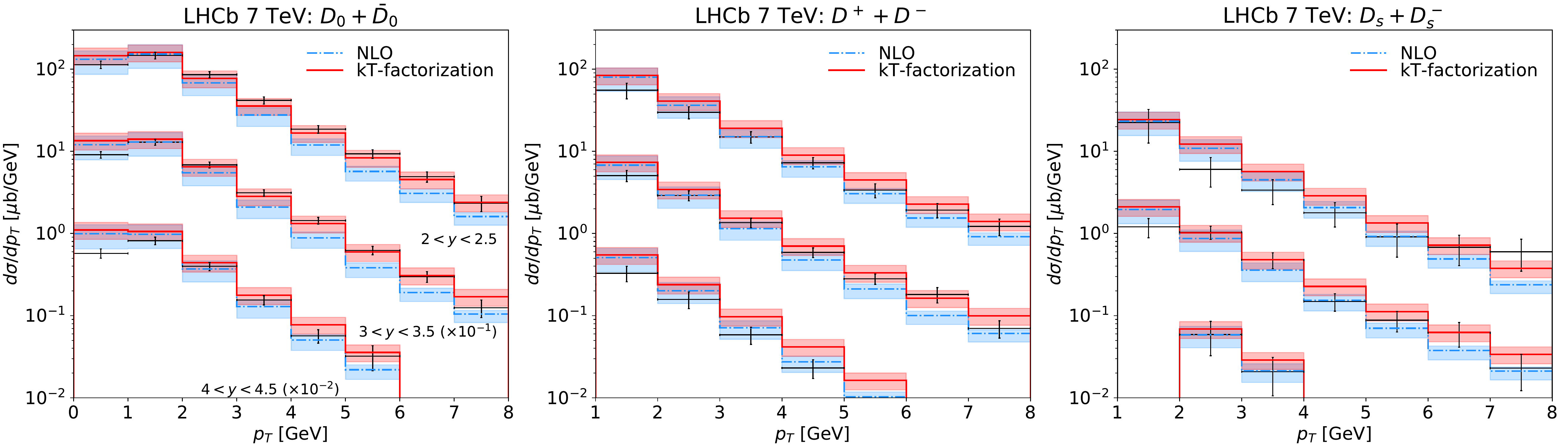}
  \caption{\label{fig:results7}\textbf{Comparison of Charm Hadron Distribution
  at 7 TeV:} Transverse momentum distributions for all three charmed mesons in
comparison to 7~TeV LHCb data using the same collinear factorization at NLO and
$k_T$-factorization setups as in \cref{fig:results1}.}
\end{figure}
%-------------------------

%-------------------------
\begin{figure}[t]
  \centering
  \includegraphics[width=0.99\textwidth]{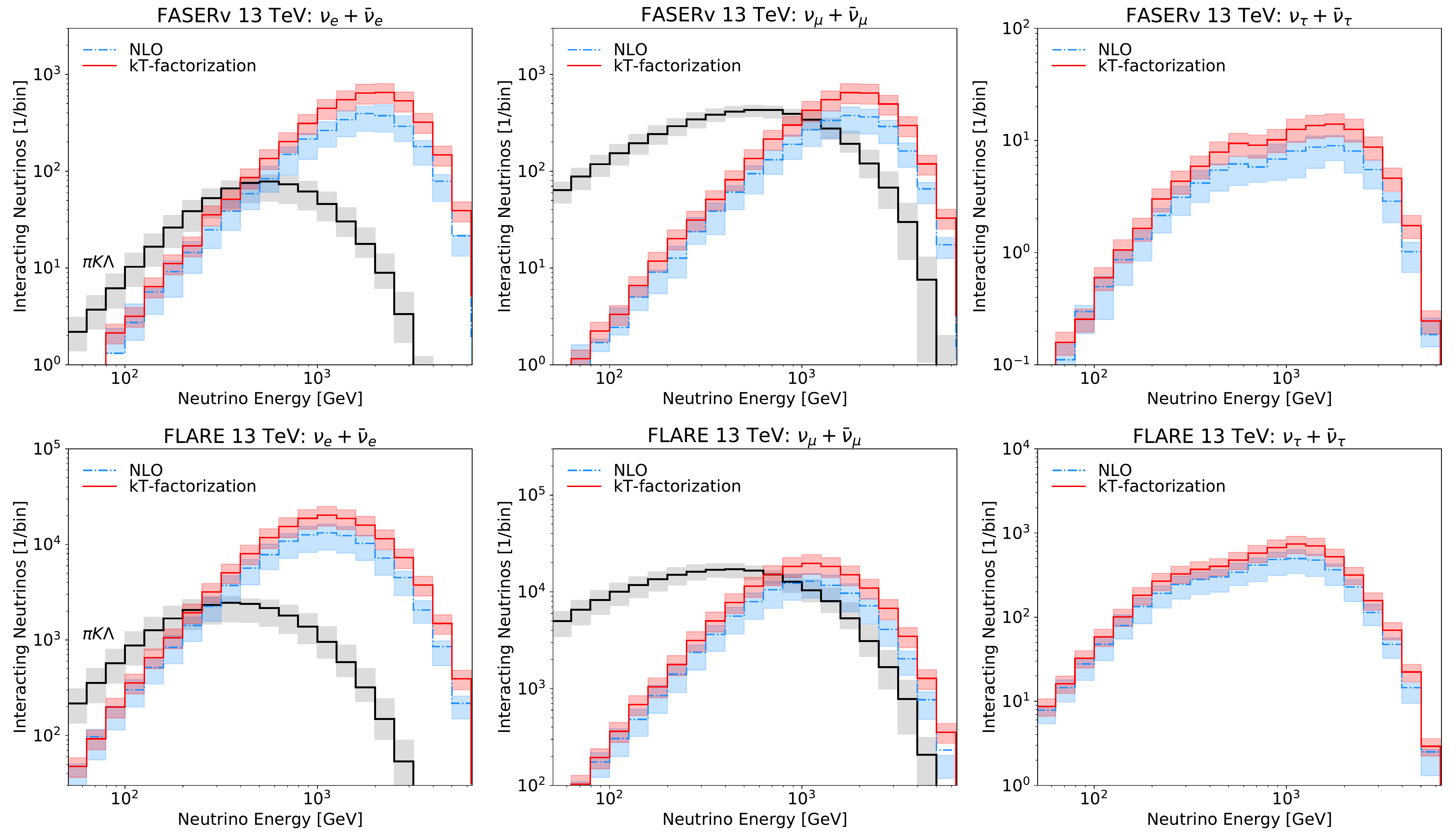}
  \caption{\label{fig:results2}\textbf{Comparison of Forward Neutrino
  Distributions:} Forward neutrino flux predictions using the same collinear
factorization at NLO and $k_T$-factorization setups as in \cref{fig:results1}.
The energy spectra of neutrinos interacting in FASER$\nu$ at Run~3 of the LHC
are shown in the top row for all three neutrino flavors. Similar distributions
for proposed FLARE detector at FPF during the HL-LHC era are shown in the
bottom row.}
\end{figure}
%-------------------------

In \cref{fig:results1}, we compare the corresponding distributions from both
approaches. We note that $k_T$-factorization with saturation gives slightly
better description of the $p_T$ shape of the LHCb  data than the NLO case.
However, we again remind the reader, that this has to be taken with caution
since this calculation includes the fitted $k$-factor which is not needed for
the NLO collinear approach.   We find that both approaches give good
description of $D^0 + \bar{D^0}$ data but when compared with LHCb data for $D^+
+ D^-$, and for $D_s + D_s^-$,  the low $p_T$ region is overestimated.  We show
distributions as a function of rapidity for different $p_T$ regions, and we
find that NLO and $k_T$-factorization with saturation give similar values for
central rapidity, but they differ at large rapidity, by about a factor of $2$,
especially for $0 < p_T <0.5 $ region.  For large values of $p_T$, this
difference is reduced. 

In \cref{fig:results7}, we also show comparison of both approaches with the
LHCb data at 7 TeV, and the description is very good in both cases. It should
be stressed that the used scales for the NLO calculation and the $k$-factor for
$k_T$-factorization at $7$ TeV are the same as extracted from $13$ TeV data. As
mentioned previously, this is encouraging since it means that the energy
dependence of the data, which is driven mainly by the $x$ evolution of the
unintegrated gluon density is captured correctly. The latter one has been taken
from the resummed approaches~\cite{Kwiecinski:1997ee, Ciafaloni:2003ek,
Ciafaloni:2003rd} which aim to reproduce both small-$x$ and collinear dynamics. 

The neutrino flux obtained using both QCD approaches is presented in
\cref{fig:results2}. The upper row shows the number of interacting neutrinos in
FASER$\nu$ operating during LHC Run~3 with an integrated luminosity of
$150~\ifb$ while the bottom row shows the neutrino events rate at FLARE at the
FPF during the HL-LHC with a luminosity of $3~\iab$. The three columns
correspond to the three neutrino flavors. The shape of the neutrino flux
remains similar for all neutrino flavors in both approaches, with the NLO
contribution slightly lower than that of the $k_T$-factorization. However, the
two approaches are very close and fall within the range of uncertainty, which
is approximately a factor of two. The black lines represent the contribution to
the neutrino flux from decays of light hadrons. Notably, we find that the
dominant contribution to neutrinos occurs above $500~\gev$ for $\nu_e$ and
above $1~\tev$ for $\nu_\mu$. Detecting $\nu_\tau$ would serve as a direct test
of charm production, as there is no contribution from pions and kaons decays.

Based on our calculation, we predict that FASER$\nu$ during LHC Run~3 is
expected to observe approximately $4000~\nu_e$, $4000~\nu_\mu$, and
$120~\nu_\tau$ charge current interactions originating from decays of charm
hadrons. The FPF, proposed to house larger neutrino detectors during the HL-LHC
era, aims to record a significantly larger sample of neutrino interaction
events~\cite{Anchordoqui:2021ghd, Feng:2022inv}. Specifically, we consider the
FLARE detector housed within the FPF, for which we assume a $1~\m \times 1~\m
\times 7~\m$ liquid argon target~\cite{Feng:2022inv}. We can see that FLARE
will detect approximately $1.4\times10^5~\nu_e$, $1.4\times10^5~\nu_\mu$, and
$6000~\nu_\tau$ from charm hadron decays. This substantial increase in
statistics will enable FPF experiments to conduct more detailed tests on
forward charm production and provide the necessary data to distinguish between
different predictions.

%%%%%%%%%%%%%%%%%%%%%%%%%%%%%%%%%%%%%%%%%%%%%%%%%%%
\section{\label{sec:conclusion}Conclusion}
%%%%%%%%%%%%%%%%%%%%%%%%%%%%%%%%%%%%%%%%%%%%%%%%%%%
Forward charm production at hadron colliders has long been recognized as an
sensitive tool for probing the strong interaction. However, until recently, it
has remained beyond the reach of the existing LHC experiments. This situation
is now changing with the start of operation of the FASER$\nu$ and SND@LHC
experiments, which are strategically positioned in the far-forward direction of
the LHC and specifically designed to detect collider neutrinos. Many of these
neutrinos originate from the decay of charm hadrons, presenting a unique
opportunity to investigate forward charm production. Together, FASER$\nu$ and
SND@LHC are projected to observe approximately ten thousand neutrinos during
the LHC's Run 3, spanning from 2022 to 2025. Looking forward, a continuation of
this collider neutrino program is envisioned for the HL-LHC era from 2029 to
2042: by utilizing larger detectors situated in the FPF it will be possible to
detect millions of collider neutrinos.

In this work, we have predicted neutrino fluxes from charmed mesons in these
forward neutrino experiments.
To this end, we have modeled charm hadron production from $pp$ collisions
at 13 TeV using different QCD and hadronization models, fitting our hadron cross-sections to
charmed meson data from LHCb to ascertain the values of parameters involved
in our models.
This also allows us to determine which QCD and hadronization models are well
tailored to describing physics at the forward rapidities that will be probed
at FASER$\nu$.

When evaluating hadron cross-sections against current collider data,
we have placed particular emphasis on the hadronization models used to convert charm
cross-sections to hadronic ones.
We have discussed how current fragmentation function based
models in the literature are not especially well motivated to
describing far forward physics, because, among other things,
they omit the potential for involving beam remnants when hadronizing.
With the end goal of accurately forecasting neutrino fluxes at FASER$\nu$,
we have, instead, devised a scheme that employs the string
fragmentation model implemented in \texttt{Pythia~8},
resulting in a more realistic representation of hadronization.
This \texttt{Pythia}-based scheme naturally overcomes most of the theoretical
shortcomings of fragmentation function based models.
We also demonstrate that the use of this hadronization scheme leads to
a significantly enhanced flux of forward neutrinos compared to those obtained
using established fragmentation functions, which results from allowing the
hadronization with beam remnants. 
This underscores the importance of utilizing an accurate fragmentation modeling.
However, we also note that the topic of forward charm hadronization warrants
further theoretical investigation.

To obtain the charm cross-sections that underpin our analysis, we have
investigated two distinct QCD models
and made noteworthy improvements to each insofar as they apply to
forward physics:
\begin{inparaenum}[\itshape~a\upshape)]
\item collinear factorization, where we use factorization and renormalization
      scales as free parameters to be determined by fitting to LHCb data, as is
      typically done in the literature, but in addition apply a $k_T$ smearing
      on the charm transverse momentum in an energy conserving way; and
\item $k_T$ factorization, which is more suitable for the description of the
      forward particle production at high energy since it resums contributions
      due to the small $x$ effects in the parton density, and where we include a
      $k$-factor to account for a mismatch in the normalization against LHCb data.
\end{inparaenum}
When using the former, we find that --- no matter
the variation of scales --- the agreement of the shape of final hadron
differential cross-sections vis-à-vis LHCb data is noticeably improved by
the allowing a Gaussian smearing of the charm transverse momentum with some
mean $k_T$.
In contrast with Ref.~\cite{Bai:2020ukz}, where this $k_T$ smearing effect
has been first discussed, our analysis explicitly conserves energy when applying
the transformation by keeping the charm $z$-momentum constant and allowing its
rapidity to vary.
We find a best-fit to 13 TeV LHCb data is obtained for
$\lbrace \mu_F, \mu_R \rbrace = \lbrace 1.75, 1.25 \rbrace m_T$ alongwith
$\ktmean = 1.5~\gev$.
When using the $k_T$ factorization scheme, our central prediction incorporates the \texttt{KS}
unintegrated distribution with a non-linear evolution for the low-$x$ gluon, and
the \texttt{CT14nlo} distribution for large-$x$ gluons.
A salient feature of our analysis is that, in order to describe the LHCb data,
we introduce a constant $k$-factor of $2.32$, determined by fitting the overall
normalization to data.
The need for the inclusion of the normalization $k$-factor in
$k_T$-factorization approach likely stems from the fact that the off-shell
partonic cross section for production of heavy quarks is only available at
lowest order.
To theoretically ascertain the value of the $k$-factor, higher orders of the
off-shell partonic cross section will need to be computed, and possibly resummed. 
We further examined the impact of systematically varying the underlying
QCD parameters, such as scale selection and parton distribution
function choices at low and high $x$ on these predictions.

We note that both QCD approaches provide a good description of the LHCb data
when paired with the \texttt{PYTHIA} hadronization scheme.
They exhibit similar energy dependence in the neutrino flux with slightly different
overall normalizations, which can be attributed to the specific QCD parameter choices.
More theoretical work with respect to the underlying uncertainties relevant to each
model is needed to improve the precision of these calculations (in particular the
small $x$ approach), as well as further experimental input in order to distinguish
between NLO collinear and $k_T$-factorization, and especially to see an onset of
parton saturation.

Using our best-fit QCD models, we have shown predictions
for neutrino fluxes of all three flavors for the ongoing FASER experiment as well
as the proposed FLARE detector at the FPF and compared them against those from the
decay of lighter mesons.
We find that, depending on the choice of QCD scheme, the electron neutrino flux
from charmed mesons dominates over those from pions and kaons starting at neutrino
energies between 400 and 500~GeV.
Furthermore, muon neutrinos from charmed meson decay become comparable to those
from pion and kaon decays at energies above 1~TeV for both QCD approaches.
Tau neutrinos, produced exclusively from heavy meson decays, provide a
background-free channel to investigate heavy meson QCD.
Our models predict between 4000 and 6000 charged current tau neutrino
interactions at FLARE with energies around 1~TeV, depending on whether one uses the
collinear NLO scheme or the $k_T$-factorization scheme respectively.

The first observation of collider neutrinos at FASER~\cite{FASER:2023zcr}
heralds the opening of a new frontier towards significantly improved
understanding of forward QCD. Further measurements at both FASER and SND@LHC
will provide a unique opportunity to gather valuable information about
small-$x$ QCD, validity of $k_T$-factorization and NLO collinear approach, and
validity of different hadronic fragmentation scenarios at forward rapidities.
In the future, the planned experiments at the FPF, with significantly improved
statistics, will become the ideal place to unravel these most important facets
of QCD. In addition, we expect that measurements of the forward neutrino
production at the LHC will provide valuable inputs for estimating the prompt
neutrino flux, reducing its theoretical uncertainties and thus providing a
better understanding of the main background for the detection of ultra-high
energy neutrinos be it from extragalactic astrophysical sources or from beyond
standard model physics including dark matter decays and annihilation.

%***********************************
\acknowledgments%
%***********************************

We thank   
    Akitaga Ariga, 
    Weidong Bai, 
    Luca Buonocore,
    Bhavesh Chauhan, 
    Rikard Enberg,
    Anatoli Fedynitch, 
    Jonathan Feng, 
    Sean Flemming,
    Rhorry Gauld,
    Yu Seon Jeong,
    Rafal Maciu\l{}a, 
    Mary Hall Reno,
    Luca Rottoli,
    Torbjörn Sjöstrand, and 
    Antoni Szczurek 
for many fruitful discussions. We are grateful to the authors and maintainers of many open-source software packages, including 
\texttt{RIVET}~\cite{Buckley:2010ar} and
\texttt{scikit-hep}~\cite{Rodrigues:2019nct}. 
A.B. acknowledges support from the Fonds de la Recherche Scientifique-FNRS, Belgium during the period this work was in progress.
F.K. acknowledges support by the Deutsche Forschungsgemeinschaft under Germany's Excellence Strategy - EXC 2121 Quantum Universe - 390833306. 
I.S. is supported by the U. S. Department of Energy Grant DE-FG02-13ER41976/sc-0009913. 
A.M.S is supported  by the U.S. Department of Energy Grant DE-SC-0002145 and  within the framework of the of the Saturated Glue (SURGE) Topical Theory Collaboration as well as by Polish NCN  Grant No. 2019/33/B/ST2/02588. 

\appendix
\section{\label{sec:appendixa}Variation of parameters in the collinear factorization}

To determine the correct global best-fits for these scales, one ought to use
all the available data for charm production in $pp$ collisions and determine
the cross-section that gives the least value of $\Delta \chisq$. However, for
our specific study where predictions for the forward neutrino flux are the end
goal, we need cross-sections that accurately describe the data at high energies
$\sqrt{s} \sim 13$~TeV and high rapidities. The highest rapidity
$d^2\sigma/dydp_T$ data at 13 TeV come from charmed meson cross-sections
observed at LHCb~\cite{LHCb:2015swx}. These include cross-sections for $D^0,
D^\pm, \text{ and } D_s$ at rapidities between $2 \leqslant y \leqslant 4.5$
binned by $0.5$, i.e.\ five bins in $y$ for each meson. We focus on this subset
of collider data to determine the scales, $\mu_R$ and $\mu_F$, that best
describes it. 

To compare our theoretical cross-sections against charmed meson cross-sections
from LHCb, we first compute the double differential cross-section
$d^2\sigma_{c\bar{c}}/dy dp_T$ for bare $c\bar{c}$ pair production using
specific values for the fragmentation and renormalization scales. We then
assume a specific fragmentation scheme, without any additional free parameters,
to hadronize the charm quarks into hadrons. The resulting differential
cross-section distribution at this stage may now be compared against
corresponding LHCb data for a measure of its goodness of fit, which we achieve
by means of a simple \chisq\ analysis. Since accurately forecasting high
rapidity cross-sections is critical towards obtaining predictions for forward
experiments like FASER, it becomes important to ensure that the goodness of fit
analysis is not skewed by the availability of significantly more data at LHCb's
lower rapidities $2 \leqslant y \leqslant 3$ rather at, say, $y \geqslant 4$.
We therefore use a $\chisq$ measure that is normalized to the number of $p_T$
bins with cross-section measurements for each rapidity bin in the LHCb data,
ensuring that each bin carries equal weight towards the measure.

Repeating this procedure for multiple $\left( \mu_R, \mu_F \right)$ values, we
generate a range of cross-sections and, for a given fragmentation scheme, we
ascertain the best-fit value of these parameters as the one that minimizes the
\chisq/d.o.f. Likewise, we also obtain the parameters corresponding to a
$1\sigma$ region of variation around the best-fit cross-section. The gives us a
set of best-fit $\left( \mu_R, \mu_F \right)$ parameters for each choice of
fragmentation scheme.

\section{\label{sec:app-nlo-ktsmear}$k_T$ smearing in collinear factorization}

%-------------------------
\begin{figure}[b]
  \centering
  \includegraphics[width=0.99\textwidth]{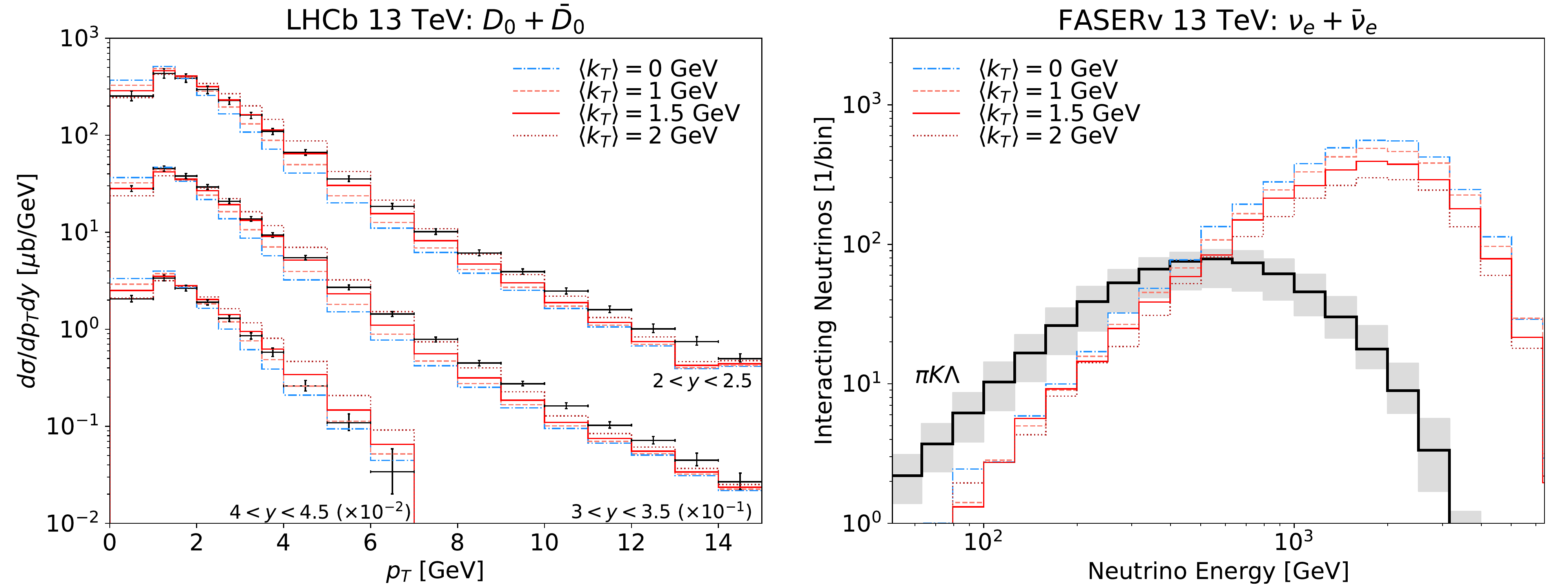}
  \caption{\label{fig:nlo_kt}\textbf{ Smearing  of ${\bf k}_T$ in Collinear Factorization:}
  Predictions using collinear factorization at NLO including $k_T$ smearing for
different values of $\langle k_T \rangle$. All predictions use fixed scales
$(\mu_F, \mu_R)=(1.75, 1.25)~m_T$, the \texttt{CT14nlo} parton distribution
function and \texttt{Pythia} with the QCD-inspired color reconnection scheme to
model fragmentation. See the main text for a detailed discussion.}
\end{figure}
%-------------------------

In \cref{fig:nlo_kt}, we present our results when applying the smearing with
different values of $\langle k_T \rangle$ to our central prediction. In this
case, when using $\langle k_T \rangle = 1.5~\gev$, our fitting routine leads to
a best-fit $(\mu_F,\mu_R) = (1.75,1.25)~m_T$. As shown in the left panel, the
shape of the transverse momentum distributions in the LHCb range changes:
events are shifted from the lowest $p_T$ bins towards intermediate $p_T$ bins.
This effect becomes stronger with increasing $\langle k_T \rangle$, and by
scanning fits made using different fixed values for $\langle k_T \rangle = 0,
0.5, 1.0, 1.5, 2.0\ldots$, we find that the best agreement with data is
obtained for $\langle k_T \rangle=1.5~\gev$. As shown in the right panel of
\figref{nlo_kt},
the corresponding neutrino fluxes are not highly sensitive to the choice of
$\langle k_T \rangle$.

\section{\label{appendixb}Variation of parameters in the $k_T$ factorization calculation}

In this appendix we include the tests of the $k_T$ factorization approach while
varying the   large $x$ gluon distribution, the order of the running coupling
and the choice of the scales.

In \cref{fig:kt_highx}, we explore the impact of the choice of integrated gluon
distributions used for the high-$x$ region, while keeping the unintegrated PDF
with saturation for the low-$x$ gluon fixed. We consider four different choices
consisting of the \texttt{CT14} and \texttt{NNPDF30} distributions at both
leading and next-to-leading order. Our results show that the next-to-leading
order distributions provide a somewhat better description of the LHCb data. In
contrast, the leading order distributions tend to overestimate the production
rate at low $p_T$, leading to an increased neutrino flux. We observe small
variations within the same order of distributions, with a corresponding
uncertainty of about $20-25~\%$ at the peak of the flux.  Therefore we choose
the NLO PDFs in the calculation as our standard choice and for a better
accuracy.

Next, in \cref{fig:kt_alphas} we study the dependence of the
$k_T$-factorization calculation on the choice of the order at which the strong
coupling is taken as well as the value of $\Lambda_{QCD}$. Our standard choice
is denoted by the `hybrid' in \cref{fig:kt_alphas}. As mentioned previously
this choice amounts to taking one power of the strong coupling in the leading
order. This is consistent with the choice used in the fit used to extract the
unintegrated \texttt{KS} gluon density in~\cite{Kutak:2012rf}. The second power
of the strong coupling is taken consistent with the choice of the large-$x$
gluon PDF, in this case \texttt{CT14nlo} set. We also compare this with two
other choices, one in which both powers of $\alpha_s$ are taken at leading
order and one in which they are taken at NLO from \texttt{CT14nlo}, labeled as
LO and NLO in \cref{fig:kt_alphas}, respectively. We see that these different
choices give a moderate spread in the predictions. We remind here that we are
using the same $k$-factor for all of these predictions, in order to isolate the
dependence on the coupling choice. The shape in $p_T$ is affected only
modestly, mainly in the low $p_T$ region. In the right panel in
\cref{fig:kt_alphas} we again show the spread of about factor of order $2$ in
the neutrino flux predictions. 

Finally, we study the dependence on the variation of the scale in the large-$x$
PDF and in the argument of the strong coupling. We vary the scale in the region
$(0.5,2.0) \langle p  _T^2 \rangle $ where we define $\langle p_T^2 \rangle  =
(p_{T1}^2+p_{T2}^2)/2$, and $p_{T1},p_{T2}$ are the transverse momenta of the
produced quark and antiquark. The results are demonstrated in
\cref{fig:kt_scales}. We observe that the variation of scales has very little
impact on both the $p_T$ dependent cross section at LHCb as well as on the
neutrino results at FASER$\nu$.

%-------------------------
\begin{figure}[t]
  \centering
  \includegraphics[width=0.99\textwidth]{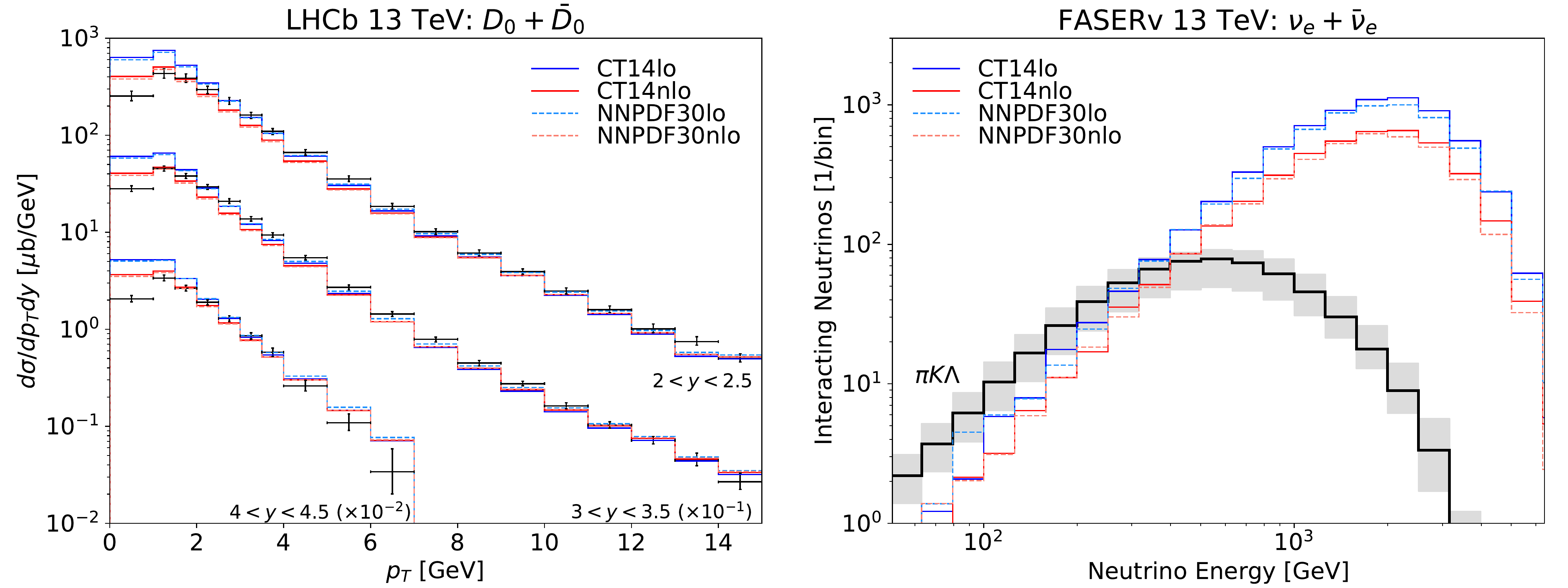}
  \caption{\label{fig:kt_highx}\textbf{High-$x$ Gluon Distribution in
  $k_T$-Factorization:} Predictions using $k_T$-factorization based on
different parton distribution of the high-$x$ gluon. All predictions use a
constant $k$-factor of $2.32$, the \texttt{KS (non-linear)} unintegrated
distribution for the low-$x$ gluon and  \texttt{Pythia} with the QCD-inspired
color reconnection scheme to model fragmentation. See the main text for a
detailed discussion.}
\end{figure}
%-------------------------

%-------------------------
\begin{figure}[t]
  \centering
  \includegraphics[width=0.99\textwidth]{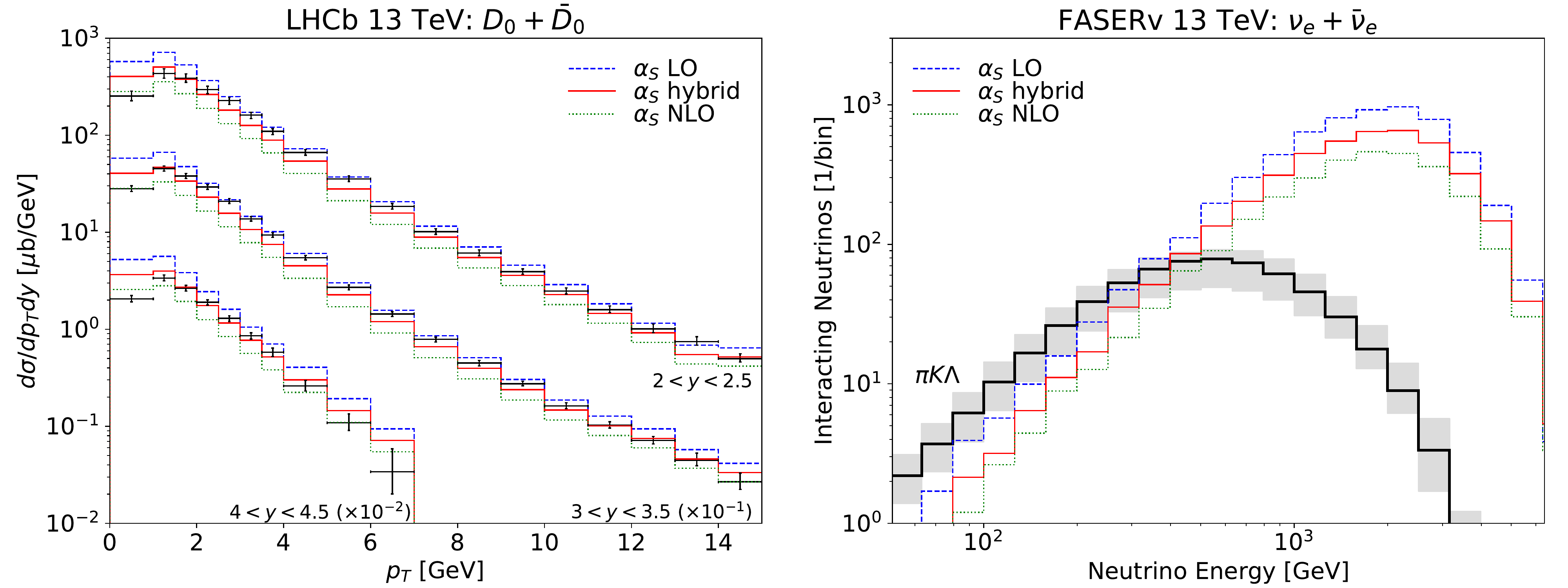}
  \caption{\label{fig:kt_alphas}\textbf{Strong Coupling in
  $k_T$-Factorization:} Predictions using $k_T$-factorization using different
choices of the strong coupling: `hybrid' (one power of coupling at LO and one
power at NLO ), `LO' (both powers of strong coupling at LO) and `NLO' (both
powers of strong coupling at NLO corresponding to the large-$x$ PDF). All
predictions use a constant $k$-factor of $2.32$, the \texttt{KS (non-linear)}
unintegrated distribution for the low-$x$ gluon, \texttt{CT14nlo} for the
high-$x$ gluon and \texttt{Pythia} with the QCD-inspired color reconnection
scheme to model fragmentation. See the main text for a detailed discussion.}
\end{figure}
%-------------------------
%-------------------------
\begin{figure}[t]
  \centering
  \includegraphics[width=0.99\textwidth]{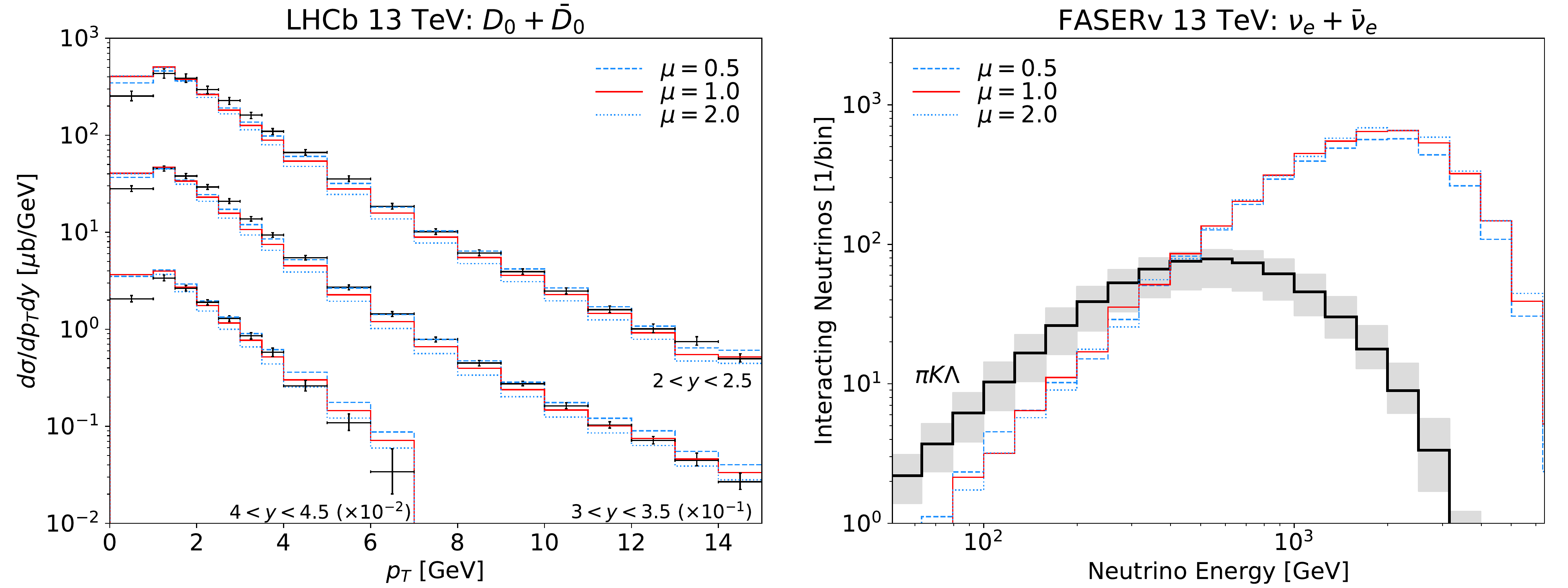}
  \caption{\label{fig:kt_scales}\textbf{Scale Choice in $k_T$-Factorization:}
  Predictions using $k_T$-factorization using scale choice of  $\mu\, \times \,
\langle p_T^2 \rangle$, with $\langle p_T^2 \rangle = (p_{T1}^2+p_{T2}^2)/2$
and values of $\mu=0.5,1.0,2.0$ for dashed-blue, solid-red and dotted-blue
curve respectively. All predictions use a constant $k$-factor of $2.32$, the
\texttt{KS (non-linear)} unintegrated distribution for the low-$x$ gluon,
\texttt{CT14nlo} for the high-$x$ gluon and \texttt{Pythia} with the
QCD-inspired color reconnection scheme to model fragmentation. See the main
text for a detailed discussion.}
\end{figure}
%-------------------------

\bibliography{references}

\end{document}